\documentclass[%
preprint,
preprintnumbers,
nofootinbib,
 amsmath,amssymb,
 aps,prd,
]{revtex4}

\usepackage{graphicx}% Include figure files
\usepackage{dcolumn}% Align table columns on decimal point
\usepackage{bm}% bold math
\usepackage{mathrsfs}
\usepackage{mathtools}
\usepackage{subcaption}
\usepackage{dsfont}
\usepackage{float}
\usepackage[normalem]{ulem}
\usepackage{ellipsis}

\usepackage{hyperref}% add hypertext capabilities
\DeclareMathOperator{\erf}{erf}

\usepackage{tikz}

\begin{document}

\title{Shifting the neutrino fog: studying the Isospin-violating Dark Matter case}

\author{Laura Duque}
\email{laura.duque@cinvestav.mx}
\affiliation{Departamento de Física, Centro de Investigación y de
Estudios Avanzados del Instituto Politécnico Nacional
Apartado Postal 14-740, 07000 Ciudad de México, México}

\author{J. M. Lamprea}
\email{jorge.lamprea@cinvestav.mx}
\affiliation{Departamento de Física, Centro de Investigación y de
Estudios Avanzados del Instituto Politécnico Nacional
Apartado Postal 14-740, 07000 Ciudad de México, México}

\author{O. G. Miranda}
\email{omar.miranda@cinvestav.mx}
\affiliation{Departamento de Física, Centro de Investigación y de
Estudios Avanzados del Instituto Politécnico Nacional
Apartado Postal 14-740, 07000 Ciudad de México, México}

\date{\today}

\begin{abstract}
First observation of solar neutrinos through coherent elastic neutrino-nucleus scattering by dark matter (DM) direct detection (DD) experiments makes the study of the neutrino fog of the most relevance. This irreducible neutrino background depends on the target material as well as other experimental parameters. Recently, it has also been remarked the dependence of the neutrino fog on the DM models under consideration. In this work, we study the case of Isospin-violating dark matter (IVDM) models, discussing specific examples of DM models and making a detailed analysis of the implications of IVDM on the neutrino fog. We also explore the conditions under which this background can be mitigated for specific DM models.
\end{abstract}

\maketitle

% ----------- introduction.tex -----------

\section{\label{sec:Introduction}Introduction}
  
Dark matter (DM) is one of the most significant and well-established mysteries in modern physics, with Weakly interacting massive particles (WIMPs) among the most promising candidates. WIMPs naturally appear in many theories beyond the Standard Model and also naturally freeze-out in the early universe, with an expected relic abundance matching the observed DM density. This cosmological coincidence is behind the motivation for a wide variety of experimental searches aimed at directly detecting WIMPs. DM direct detectors are sensitive to the tiny nuclear recoils that are expected if a WIMP from the galactic halo scatters-off a target nucleus. They are usually located deep underground to shield them from cosmic-ray backgrounds. DM direct detection (DD) experiments have shown an accelerated progress both in the size of the target as well as in detection techniques. They have increased from kilogram-scale detectors to current multi-ton-scale detectors, with liquid noble gas time projection chambers being the preferred technique, as seen in XENON1T~\cite{XENON1T}, LUX~\cite{Akerib_2016}, and PandaX~\cite{PandaX}. The next generation of experiments, including LZ~\cite{LZ} and XENONnT~\cite{XENONnt}, will improve the sensitivity in the parameter space.

As DM DD experiments grow in size and improve the low-energy thresholds, they become sensitive to the irreducible background of astrophysical neutrinos. More specifically, Boron and hep solar neutrinos can interact with the detector.  
These neutrinos could scatter-off electrons (E$\nu$Es), but the most important background for WIMP searches arises from Coherent Elastic Neutrino-Nucleus Scattering (CE$\nu$NS). This is a process happening at low momentum transfer in which neutrinos scatters-off an entire nucleus, having a large cross-section that scales with the square of the number of neutrons~\cite{Abdullah:2022zue}. This process has been theorised for a long time~\cite{Freedman} and was confirmed experimentally by the COHERENT collaboration~\cite{Akimov_2017}. The CE$\nu$NS interaction produces a low-energy nuclear recoil, which is the exact signal that DM DD experiments are designed to detect. 

The CE$\nu$NS background imposes a fundamental limitation on DD experiments, which is often referred to as the ``neutrino floor"~\cite{Cowan:2010js, Billard, Billard:2011zj}. In the neutrino floor region, the expected neutrino background almost perfectly imitates the expected WIMP signal. For example, ${}^8$B solar neutrinos produce a recoil spectrum nearly identical to that of a $6$~GeV WIMP with a DM-nucleon cross section of $5\times 10^{-45}$~cm$^2$, while atmospheric neutrinos mimic a $100$~GeV WIMP with a DM-nucleon cross section of $2.5 \times 10^{-49}$~cm$^2$ (See Section~\ref{sec:Neutrinos})\footnote{ the analog neutrino fog for DM-electron scattering, the reader can look at Refs.~\cite{Carew:2023qrj,Essig:2018tss}, while for Migdal effect, please see Refs.~\cite{Maity:2024hzb,Herrera:2023xun}}. However, the neutrino floor does not represent a hard ``floor", but rather a limit that can change depending on the target material and the DM parameters. That is why the term ``neutrino fog"~\cite{O_Hare_2021} is more appropriate, since the limit is not an absolute barrier, but rather a transitional region where identifying a recoil as a DM signal becomes ambiguous due to the irreducible neutrino background. This ``neutrino fog" has been reached experimentally in recent years, first in 2024 by the PandaX-4T~\cite{PandaxSolar} and XENONnT~\cite{XenonntSolar} collaborations, who reported the first indications of ${}^8$B neutrinos at 2.64$\sigma$ and 2.73$\sigma$ significance, respectively. After these initial indications, the LZ experiment~\cite{LZSolar} succeeded in making a measurement with a statistical significance of 4.5$\sigma$ in late 2025. This recent CE$\nu$NS measurement in DD experiments makes more urgent to have a clear understanding of the neutrino fog shape and its robustness against different DM models.

On the other hand, to compute the expected DM signal in DD experiments, we usually rely on the hypothesis that the DM candidate interacts with equal strength with the protons and neutrons of the target material. Such hypothesis has important consequences on the expectations for the neutrino fog; therefore, it is important to question how robust the Isospin conservation hypothesis is.
Although this could be a reasonable first guess, there is no reason why DM must be blind to the different Isospins of these nucleons, and some works have analysed their impact in DD experiments~\cite{Catena:2015uua,Brenner:2022qku,AvisKozar:2023iyb}. Actually, in the Standard Model (SM), this statement does not hold, and, for instance, the neutrino coupling to protons in CE$\nu$NS is much more smaller than the one to neutrons.  Although there is no strong reason why DM should present the same isospin violation as in the Standard Model, there is also no reason why there should have isospin conservation. 

Based on this idea, we have studied in this article the impact of Isospin-violating dark matter (IVDM) on predictions for the neutrino fog. IVDM proposes that DM interacts with protons ($f_p$) and neutrons ($f_n$) with different strengths. This framework relaxes the standard assumption of DM models where these couplings are identical. Considering IVDM introduces an asymmetry in the DM couplings to protons and neutrons and may induce a destructive interference for a given atomic nucleus. This change in the neutron-to-proton coupling ($f_n \neq f_p$) can significantly reduce the expected event rate for a given target material.  In a previous work, the “xenophobic” scenario, typically characterised by $f_n/f_p \approx -0.7$, has been explored~\cite{Feng_2011}, motivated by the destructive interference that made xenon negative searches consistent with the DAMA/LIBRA and CoGeNT signals. Furthermore, it is worth noting that the neutrino fog has been analysed in the context of light mediators~\cite{deromeri2025}, and the effects on this limit have also been explored for an IVDM  model, especifically a $Z'$ model~\cite{lozano2025}. In our analysis, however, we use the profile likelihood ratio test approach, which offers a more robust analysis of the neutrino fog. Unlike the isocurve method, that treats the neutrino fog as a fixed limit, this method integrates all the uncertainties as nuisance parameters, converting the neutrino fog into a dynamic limit that evolves depending on how much the DM  signal overlaps with the background neutrinos. Additionaly, other models are explored in the following sections. 

This article is organised as follows. Before describing the statistical procedure for computing the neutrino fog, we first discuss, in Section~\ref{sec:DarkMatter}, the technical details for the computation of DM events that may be expected in a DM DD experiment. We devote Section~\ref{sec:Neutrinos} to describe the corresponding computations for the neutrino case. Afterwards, we proceed with the detailed explanation for the calculation of the neutrino fog in Section~\ref{sec:NeutrinoFog}. The results for our case of interest, the IVDM, are discussed in Section~\ref{sec:IVDM}. Finally, we draw our conclusions in Section~\ref{sec:Conclusions}

% ----------- DarkMatter.tex -----------

\section{\label{sec:DarkMatter}  Dark Matter Expected Signal}
Before discussing the IVDM effects on the neutrino fog predictions, we must discuss the typical signal that is searched for in DD experiments.
DD experiments search for a DM particle interacting with the nuclei in the detector, assuming that there is a very weak interaction with the detector material. In this section, we show the signal produced by a WIMP $\chi$ with mass $m_\chi$ scattered-off a nucleus target of mass $m_N = m_A/N_A $, where $m_A$ is the molar mass and $N_A$ is the Avogadro number in units of mol${}^{-1}$. The differential rate $R_\text{DM}$ of DM events per recoil energy $E_r$ is given by~\cite{Lewin:1995rx}\\
\begin{equation}
    \frac{d R_{\mathrm{DM}}}{dE_r} = \epsilon \frac{\sigma_0 \rho_0}{2 m_{\chi} \mu_N^2 } \int_{v > v_\text{min}} \frac{f(\mathbf{v})}{v} \, d^{3}v,
\end{equation}
\\
where $\epsilon$ is the detector exposure in units of ton-year, $\rho_ 0 = 0.3~\mathrm{GeV~c^{-2}~cm^{-3}}$, the local DM mass density, $\mu_N = \frac{m_N + m_\chi }{m_\chi m_N}$ is the nucleus-DM reduced mass, $f(\mathbf{v})$ the DM velocity distribution and $\sigma_0$, the cross section, which is assumed to be independent of the velocity, $\mathbf{v}$, and the momentum transfer, $q$. The integral in the above equation is computed over all the velocities greater than the minimum WIMP velocity, $v_\text{min}$, needed  to achieve a nuclear recoil with energy $E_\text{r}$,
\begin{equation}
v_\text{min} = \sqrt{\frac{m_N E_\text{r}}{2 \mu_N}}.
\end{equation}

The cross section, $\sigma_0$, is defined as the velocity-independent and thermal averaged WIMP-nucleus cross section, which is comprised of the Spin-Independent (SI) and the Spin-Dependent (SD) part as
\begin{equation}
\sigma_0 =  (\sigma_0^\text{SI} F^2_\text{SI}(q)+ \sigma_0^\text{SD} F_\text{SD}^2(q)), 
\end{equation}
where $F_\text{SI,SD}$ are the SI and SD nuclear form factors respectively. For simplicity, we will consider that the proton and neutron form factors are equal. Moreover, in our case study, we will focus on the SI part of this cross-section and denote $F_\text{SI} (q) = F(q)$. Then, in the case where the WIMP scatters-off a nucleus of $Z$ protons and $A-Z$ neutrons, the SI cross section takes the form
\begin{equation}\label{eq:sigma}
    \sigma^\text{SI}_0 = \frac{4\mu_n^2}{\pi} \left[ Z f_p + (A-Z) f_n \right] ^2 ,
\end{equation}
where $f_p$ and $f_n$ are the proton and neutron couplings to WIMP, and $\mu_n$ the nucleon-WIMP reduced mass. \\

The above mentioned nuclear form factor, $F^2(q)$, describes the loss of coherence as the momentum transfer $q$ increases. For extremely low momentum transfer, $q \ll 1/R$, there is no sensitivity to the nuclear structure and $F(q) = 1$. When $q$ increases, the WIMP  begins to perceive the spatial distribution of nucleons inside the nucleus, causing the individual scattering amplitudes to no longer add in phase. For the nuclear form factor, in this work, we have considered the Klein-Nystrand (KN) parameterisation~\cite{Papoulias_2019}:\\
\begin{equation}
        F_{KN}(q)=3 \frac{j_1(q R_A)}{q R_A} \left[1+(q a_k)^2\right]^{-1}.
        \label{eq:FormFactor}
\end{equation}\\
In this expression, $j_1$ is the first-order spherical Bessel function, $R_A$ is the nuclear radius, which is often approximated by $R_A = 1.2 \, A^{1/3}$ fm~\cite{Klein_1999}, while $a_k = 0.7$ fm is a parameter related to the nuclear skin thickness.

Regarding the DM velocity distribution in our Galactic Halo, we  assumed it to follow a truncated Maxwellian distribution (the Standard Halo Model~\cite{Savage, Freese_2013}) with $\sigma_v$ the velocity dispersion, $v_0 = \sqrt{2/3} \, \sigma_v = 220~\text{km/s}$ the local circular velocity of the Earth with respect to the galactic centre, and $v_\text{esc} = 544~\text{km/s}$ the DM escape velocity. In the galactic centre frame of reference, this distribution takes the form
\begin{equation}
    f(\textbf{v})=\begin{cases}
    \frac{1}{N_{\mathrm{esc}}}\left(\frac{3}{2\pi\sigma_v^2}\right)^{3/2} e^{-3\mathbf{v}^2/2\sigma_v^2}, & \text{for $|\mathbf{v}|<v_{\mathrm{esc}}$}\\
    0, & \text{otherwise.}
    \end{cases}
\end{equation}\\
and the normalisation 
\begin{equation}
N_{\mathrm{esc}} = \erf(z) - \frac{2}{\sqrt{\pi}} z e^{-z^2},
\end{equation}
where 
\begin{equation}\label{eq:Z}
z  \coloneq v_\text{esc}/v_0
\end{equation}
The integral of the inverse of the velocity distribution $\eta(v_{\mathrm{min}}) =  \int_{v > v_\text{min}} \frac{f(\textbf{v})}{v}d^{3}v$ averaged over the solar cycle and boosted to the laboratory frame of reference, is analytically calculated as\\
\begin{equation}
     \eta(v_{\mathrm{min}})=\begin{cases}
         \frac{1}{v_{\mathrm{obs}}}, & \text{for $z<y$, $x<|y-z|$}\\
         \frac{1}{2N_{\mathrm{esc}}v_{\mathrm{obs}}}[\mathrm{erf}(x+y)-\mathrm{erf}(x-y)-\frac{4ye^{-z^2}}{\sqrt{\pi}}], & \text{for $z>y$, $x<|y-z|$}\\
         \frac{1}{2N_{\mathrm{esc}}v_{\mathrm{obs}}}[\mathrm{erf}(z)-\mathrm{erf}(x-y)-\frac{2}{\sqrt{\pi}}e^{-z^2}(y+z-x)], & \text{for $|y-z|<x<y+z$}\\
         0, & \text{for $y+z<x$}
         \end{cases}
\end{equation}\\
where 
\begin{equation}
    x \coloneq v_{\mathrm{min}}/v_0, \qquad
    y \coloneq v_{\mathrm{obs}}/v_0, 
  \end{equation}
and $v_\text{obs} = $ 232~\text{km/s}~\cite{Savage, Freese_2013} is the averaged laboratory velocity with respect to the DM rest frame of reference.

With this information, we can compute the expected number of WIMP events in the $i$-th energy bin in the detector 
\begin{equation}\label{eq:WIMPEvents}
N^i_\chi (m_\chi, \sigma_{\chi-n} ) =  \int_{E_i}^{E_{i+1}} \frac{d R_\text{DM}}{d E_r} d\,E_r.
\end{equation} \\
This expected number of WIMP events, for a given DM candidate mass and cross-section, will be used in Section~\ref{sec:NeutrinoFog} to calculate the neutrino fog. Besides, the expected number of neutrino CE$\nu$NS events, which is the subject of the next Section, will also be needed for such a computation. 

% ----------- Neutrinos.tex -----------

\section{\label{sec:Neutrinos}Coherent Elastic Neutrino-Nucleus Scattering Expected Signal}

\begin{figure}[htb]
\includegraphics[width=.49\linewidth]{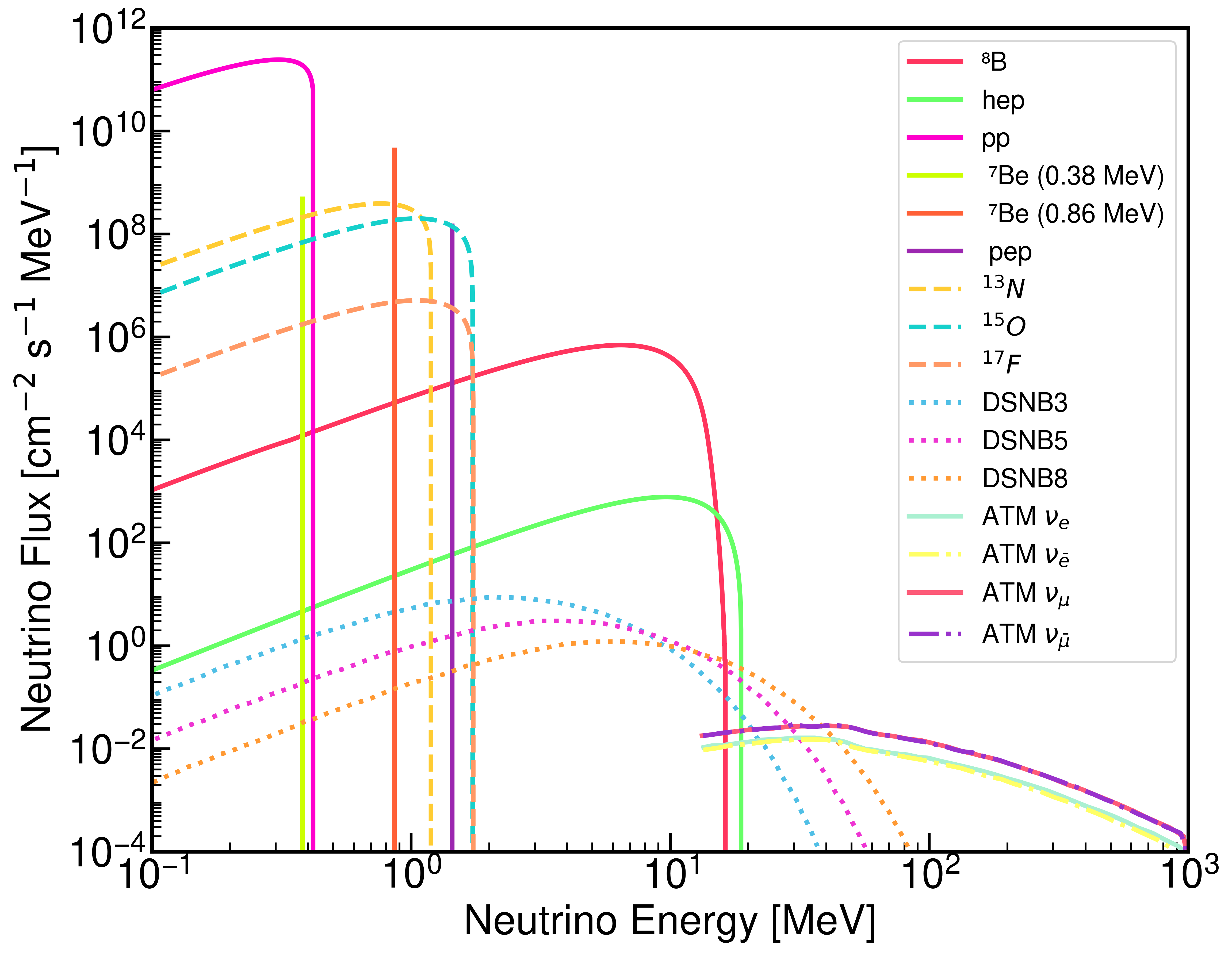}\hfill
\includegraphics[width=.49\linewidth]{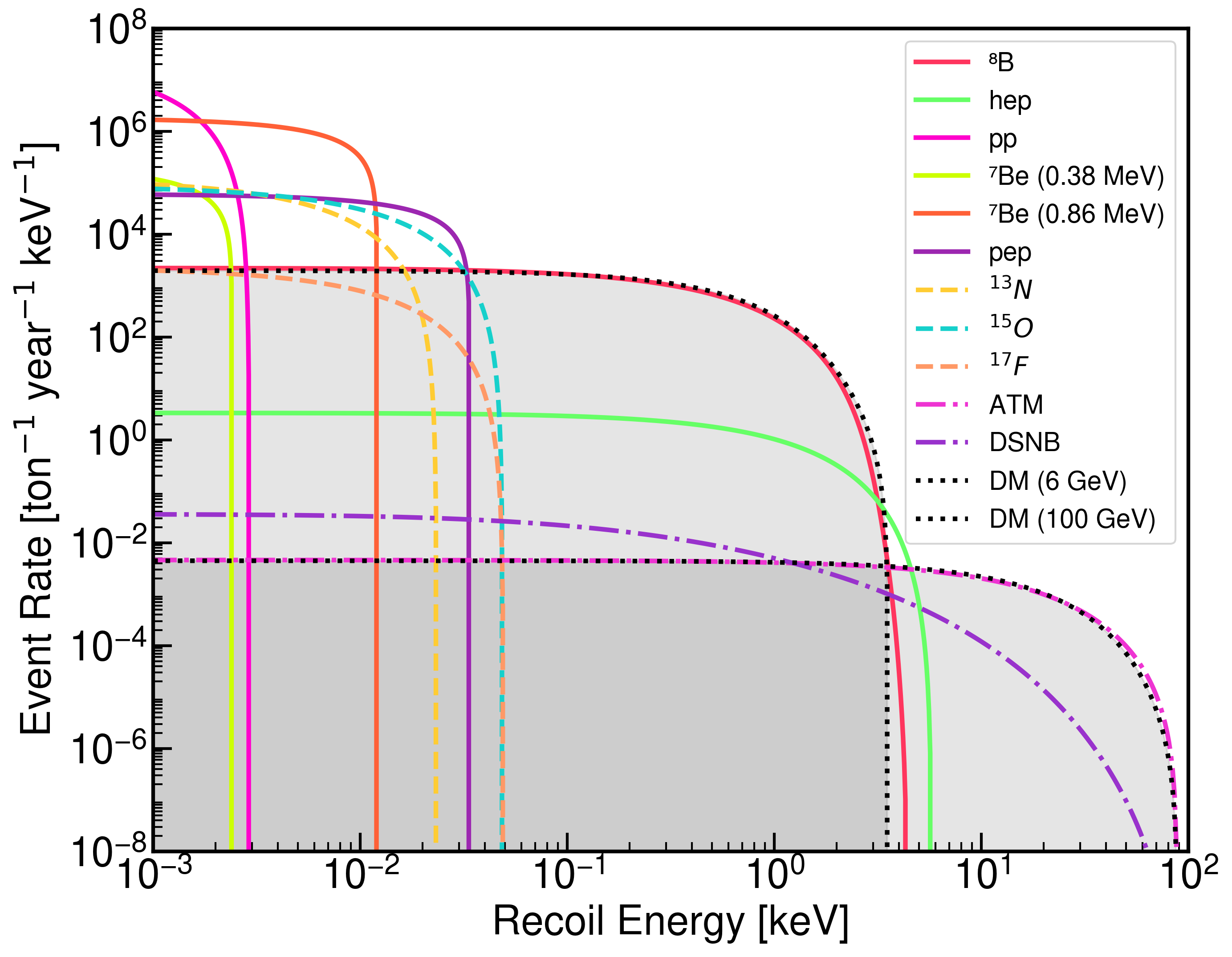}
\caption{\textbf{Left:} Neutrino energy spectra for Solar, Atmospheric (Atm), and the DSNB~\cite{Baxter_2021}. \textbf{Right:} Xenon recoil spectra showing the irreducible neutrino backgrounds (coloured) compared to potential WIMP signals (black dotted). The recoil spectra of a WIMP with mass $6$ GeV and cross section $\sigma_{\chi-n} = 5\times10^{-45}$ cm$^2$ is shown to overlap with $^8$B neutrinos, while a  $m_\chi = 100$ GeV and $\sigma_{\chi-n} = 2.5\times10^{-49}$ cm$^2$ WIMP overlaps with atmospheric neutrinos.}
\label{fig:Events}
\end{figure}

Neutrinos represent the principal irreducible background in the DD searches for DM. They interact with the detector through CE$\nu$NS, producing recoils that mimic the expected signal from a WIMP in the detector (see Fig.~\ref{fig:Events}).  The coherence in this process is guaranteed  if the momentum transfer, $q$, is small, making the neutrino wavelength ($\lambda \sim 1/q$) larger than the radius of the nucleus target, $R_A$.\\

The differential event rate for CE$\nu$NS arises from the convolution of the CE$\nu$NS differential cross section with the neutrino flux,
\begin{equation}
    \frac{dR_\nu}{dE_r} = \epsilon \frac{N_A}{m_A} \int_{E_\nu^\text{min}}^{E_\nu^\text{max}}\frac{d\Phi}{dE_\nu}\frac{d\sigma}{dE_r}dE_\nu,
\label{eq:DiffRecNeu}\end{equation}
where the minimum neutrino energy to generate a nuclear recoil with energy $E_r$ is $E_\nu^\text{min} = \sqrt{m_N E_r/2}$;  $\epsilon$ is the exposure in units of ton-year; $m_A$ is the molar mass of the target nucleus, and the differential cross section is given by \cite{DeRomeri2023}
\begin{equation}
    \frac{d\sigma}{dE_r} (E_r, E_\nu)=\frac{G_{F}^{2} m_N}{2\pi} \mathcal{Q}_{W}^2 F^2(q) \left( 2-\frac{m_N E_r}{E_{\nu}^2} \right).
    \label{eq:CEvNSCrossSection}
\end{equation}
Here, $G_F$ is the Fermi constant and $\mathcal{Q}_W$ is the vector weak charge, defined as
\begin{equation}
    \mathcal{Q}_W^2=\left( Z g_p^V + N g_n^V \right)^2,
\end{equation}
where $Z$ indicates the number of protons in the nucleus, and $N = A-Z$ the number of neutrons. The vector couplings of the proton and neutron are $g_p^V = 1/2 - 2\sin^2\theta_W$ and $g_n^V = -1/2$, respectively. It is worth mentioning that, considering the value of the weak mixing angle at low energies, $\sin^2{\theta_W} = 0.2368$~\cite{ParticleDataGroup:2024cfk, Erler:2017knj}, the proton vector coupling is very small; therefore, as it is well known, the weak charge is dominated by the neutron contribution, making CE$\nu$NS especially sensitive to the number of neutrons in the nucleus. 

Finally, the expected number of neutrino events (N$_\nu^i$) within a specific energy bin, $i$, is calculated by integrating the differential recoil rate, given in Equation~\ref{eq:DiffRecNeu}, over the energy limits for each bin 
\begin{equation}\label{eq:NeutrinoEvents}
     N_\nu^i (\Phi)= \int_{E_i}^{E_{i+1}} \frac{dR_\nu}{dE_r}\,dE_r,
\end{equation}
\\
where we have considered the contribution of all possible neutrino fluxes $\Phi = (\phi_1, \ldots, \phi_{n_\nu})$ reaching the detector, which are displayed in Table~\ref{tab:1}. Such neutrino fluxes include solar neutrinos created from thermonuclear reactions in the core of the Sun (primarily through the proton-proton (pp) chain and the CNO cycle); this process produces a spectrum of different components, from low-energy pp neutrinos to high-energy ${}^8$B neutrinos. Other sources include the atmospheric neutrinos, which are created when high-energy cosmic rays interact with the Atmosphere, resulting in a cascade of particles, including a flux of muon and electron neutrinos, and finally the DSNB~\cite{Baxter_2021},  which is comprised  of the isotropic glow of neutrinos from all past core-collapse supernovae. In Fig.~\ref{fig:Events}, we have shown the energy spectra for all of these flux components.

\begin{table}[htb]
\centering
\begin{tabular}{|c|c|c||c|c|c|}
\hline

\textbf{} & \textbf{Normalisation} [cm$^{-2}$ s$^{-1}$] & \textbf{Unc.} & \textbf{} & \textbf{Normalisation} [cm$^{-2}$ s$^{-1}$] & \textbf{Unc.} \\
\hline
pp & $5.98\times10^{10}$ & $0.6\%$ & $^{13}$N & $2.78\times10^8$ & $15\%$ \\
pep & $1.44\times10^8$ & $1\%$ & $^{15}$O & $2.05\times10^8$ & $17\%$\\
$^{7}$Be (0.861 MeV) & $4.35\times10^9$ & $3\%$ & $^{17}$F & $5.29\times 10^6$ & $20\%$ \\
$^{7}$Be (0.383 MeV)& $4.84\times10^8$ & $3\%$ & DSNB & $86$ & $50\%$ \\
$^{8}$B & $5.25\times10^6$ & $4\%$ & Atm & $10.5$ & $20\%$ \\
hep & $7.98\times10^3$ & $30\%$ &  &  &  \\
\hline
\end{tabular}
\caption{Summary of the normalisation factors and their associated uncertainties for the principal neutrino backgrounds. The sources include solar neutrinos from the pp-chain (pp, pep, $^7$Be, $^8$B, hep) and the CNO cycle ($^{13}$N, $^{15}$O, $^{17}$F), as well as atmospheric (Atm) neutrinos and the DSNB~\cite{Baxter_2021}. The normalisation factors are given in units of cm$^{-2}$ s$^{-1}$.}
\label{tab:1}
\end{table}

% ----------- NeutrinoFog.tex -----------

\section{\label{sec:NeutrinoFog} Neutrino Fog Computation}

With neutrino and DM expected rates already computed, we can turn now our attention to the computation of the expectations for the neutrino fog.
There are several ways to define the neutrino fog limit. We derive our neutrino fog definition following the already known discovery limits procedure first developed in~\cite{Cowan:2010js, Billard, Billard:2011zj}. This is defined as the limit in the cross-section versus DM mass, $\sigma_d$, at which a given experiment has a 90\% probability of detecting a DM particle with mass, $m_\chi$, and with a scattering cross section $\sigma_{n-\chi} > \sigma_d$ at 3$\sigma$ C.L.   

In order to construct our discovery limit, we define the Likelihood function as the product of the Poissonian distribution of all the bins ($N_\text{bin} = 150$) multiplied by the Gaussian distribution, ${G} (\phi^j) $, of each neutrino flux normalisation $\Phi = (\phi_1, \ldots , \phi_{n_\nu})$ in Table~\ref{tab:1},
\begin{equation}
     L (m_\chi,\sigma_{\chi-n},\Phi) = \prod_{i=1}^{N_\text{bin}} \mathcal{P} \left(N_{obs}^i | N_{exp}^i \right)  \times \prod_{j=1}^{n_\nu} G(\phi^j).
\end{equation}
We have considered $N_\text{bin} = 150$ with a threshold energy of $E_r^1 = E_\text{th} = 10^{-3}$~eV and a  maximum recoil energy, $E_r^{150} = 1$ MeV.  Let us point out that the present statistical treatment is very elaborate and helpful as we can, for instance, derive the limit in the same way in both low- and high-event-number scenarios.

The Poissonian distribution of the number of events expected $N_{exp}$, given the observed ones $N_{obs}$, is  written  as
\begin{equation}
     \mathcal{P}(N_{obs} | N_{exp}) = \frac{(N_{exp})^{N_{obs}} e^{-N_{exp}}}{N_{obs}!}.
\end{equation}
The profile likelihood ratio consists of the confrontation of two hypotheses. The background  hypothesis $H_0$ considers only the number of neutrino events (Eq.~\ref{eq:NeutrinoEvents}) as background, while the test hypothesis $H_1$ considers the neutrinos and DM events (Eq.~\ref{eq:WIMPEvents}).
Thus, in the background hypothesis, the number of expected and observed events in the bin $i$ is
\begin{equation}
     H_0 : \hspace{0.5cm} N^i_{obs} = N^i_\nu + N^i_\chi, \hspace{0.5cm} N^i_{exp} = N^i_\nu,
\end{equation}
while the test hypothesis
\begin{equation}
     H_1 :  \hspace{0.5cm} N^i_{obs} = N^i_\nu + N^i_\chi, \hspace{0.5cm} N^i_{exp}=N^i_\nu + N^i_\chi.
\end{equation}

In order to test the background-only hypothesis against the alternative hypothesis, we use the profile likelihood ratio 
\begin{equation}
     \lambda(0)=\frac{ L_0 ( m_\chi, \sigma_{\chi-n}=0, \hat{\Phi}) }{ L_1 ( m_\chi,\hat{\hat{\sigma}}_{\chi-n}, \hat{\hat{\Phi}} ) }.
\end{equation}
where the values of $\hat{\Phi}$ maximise the constrained Likelihood, $L_0$, while for the unconstrained Likelihood, $L_1$,  we have used the results of~\cite{Cowan:2010js,AristizabalSierra:2021kht}. In this sense, we utilise the Asimov dataset, where one representative dataset gives the maximum unconstrained likelihood at the mean values of $\Phi$. 

Finally, the test statistic $q_0$, in this case, is defined as 
\begin{equation}
	q_0 = 
     \begin{cases}
     -2  \log \lambda(0), & \text{$\hat{\sigma}_{\chi -n} > 0$} \\
     0, & \text{$\hat{\sigma}_{\chi-n} \le 0$}
     \end{cases}
\end{equation}
where large values of $q_0$ prefer the background hypothesis $H_0$ over the alternative one $H_1$.

The $p$-value is
\begin{equation}
p_0 = \int_{q_0^\text{obs}}^\infty f(q_0| H_0) \, d\,q_0,
\end{equation}
where $f(q_0|H_0)$ is the probability distribution of the test statistic under the background-only hypothesis. 
From Wilk's theorem~\cite{Cowan:2010js}, $q_0^\text{obs}$ asymptotically behaves like a $\chi^2$ distribution with only one degree of freedom. Thus, the significance is related as $Z = \sqrt{q_0^\text{obs}}$, where $Z \geq 3$ gives the 90\% C.L. discovery limit.

\begin{figure}%[t]
\includegraphics[width=.75\linewidth]{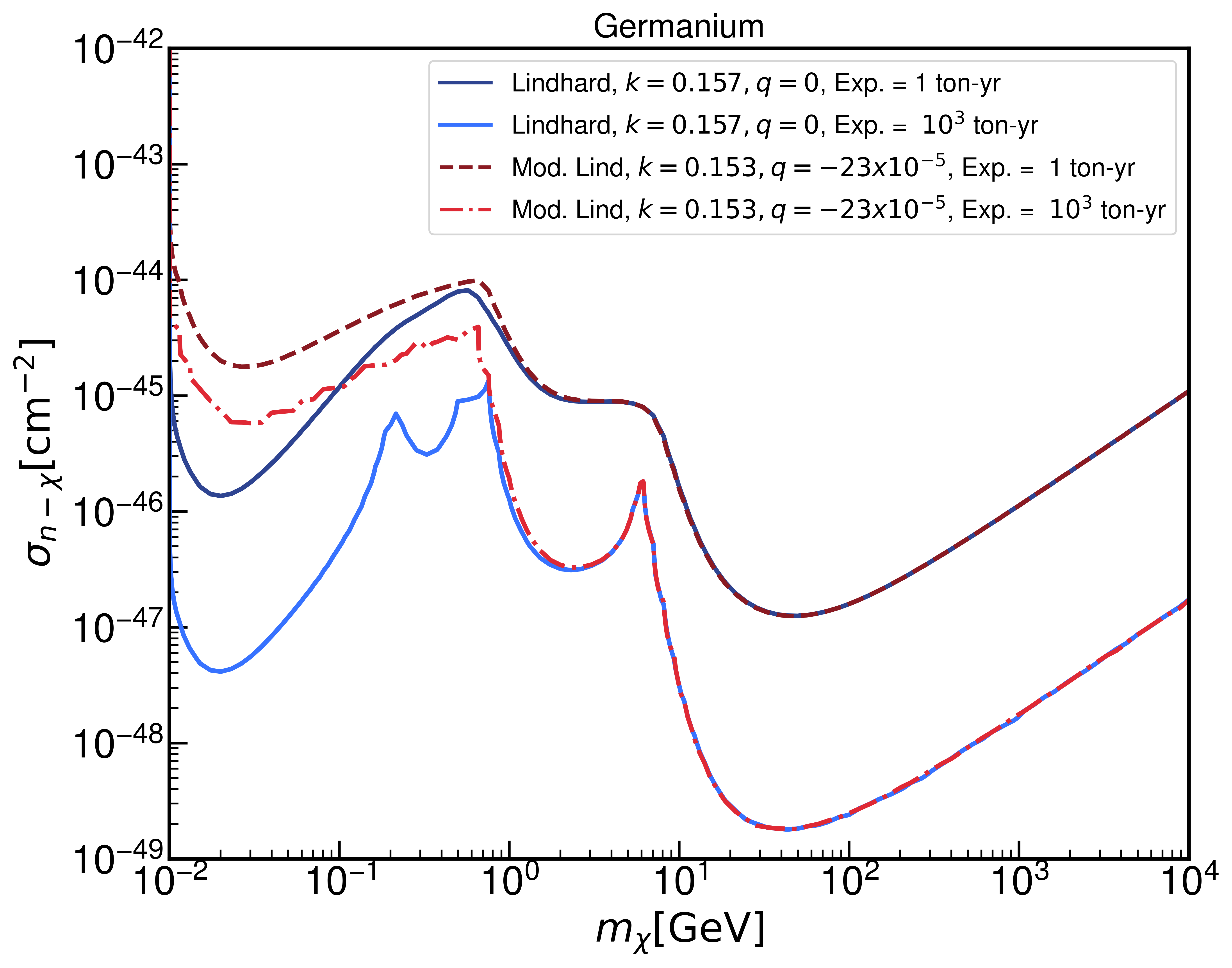}
\caption{Modification of the neutrino fog in the WIMP-nucleon SI cross-section, $\sigma_{n-\chi}$, versus WIMP mass, $m_\chi$ when we consider the impact of the Migdal effect in the Lindhard model in the case of Isospin conservation $f_n/f_p = 1$. The solid lines represent the expected neutrino fog for a germanium-type detector with a exposure of one ton-year, in dark blue, and one kton-year, in blue, in the Lindhard model~\cite{osti_4701226}, while the dashed lines describes the correction due to the Migdal effect, parametrised as in~\cite{Sorensen:2014sla} in dark red and red for the same exposures. As expected, only the low-mass region has a drastic change.}
\label{Fig:Migdal}
\end{figure}

With all the computations described up to here, we are now able to compute the neutrino fog both standard and non-standard scenarios. But before starting such a discussion it is important to notice that different systematic uncertainties can also play an important role in the determination of the neutrino fog. Our results in the next section assumes a good knowledge of such systematic effects in the experimental set ups that we have considered.

In order to illustrate the relevance of systematics in our analysis, we have chosen two different systematic uncertainties as example, they are the quenching factor uncertainties and the binning. Other systematics of relevance, such as energy resolution and specially flux normalisation should also be trated with care in any analysis. Coming back to the case of quenching factor uncertainties we have utilised the parametrisation of the Migdal effect in the Lindhard model~\cite{Sorensen:2014sla} for a germanium based detector. We can notice in Fig.~(\ref{Fig:Migdal}), that the Migdal effect can affect the low recoil energy quenching prediction  $E_r \lesssim 5~\text{keV}_{nr}$ and therefore the low-mass regime, $m_\chi \lesssim 1$~GeV. Such effect would change the prediction for the quenching factor at low-energies, shifting the threshold energy prediction as well as the number of events for the low-energy nuclear recoils.

\begin{figure}%[t]
\includegraphics[width=.49\linewidth]{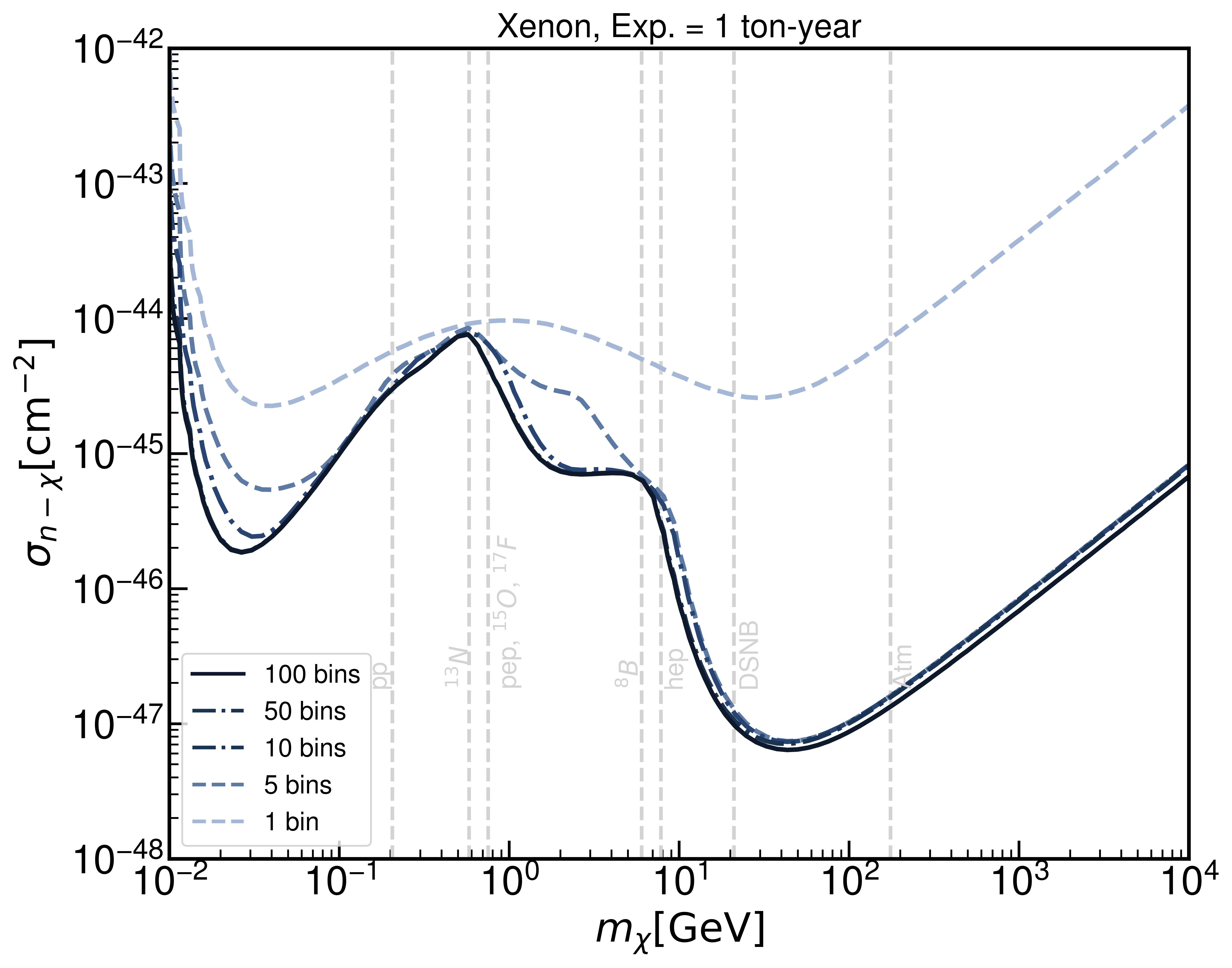}\hfill
\includegraphics[width=.49\linewidth]{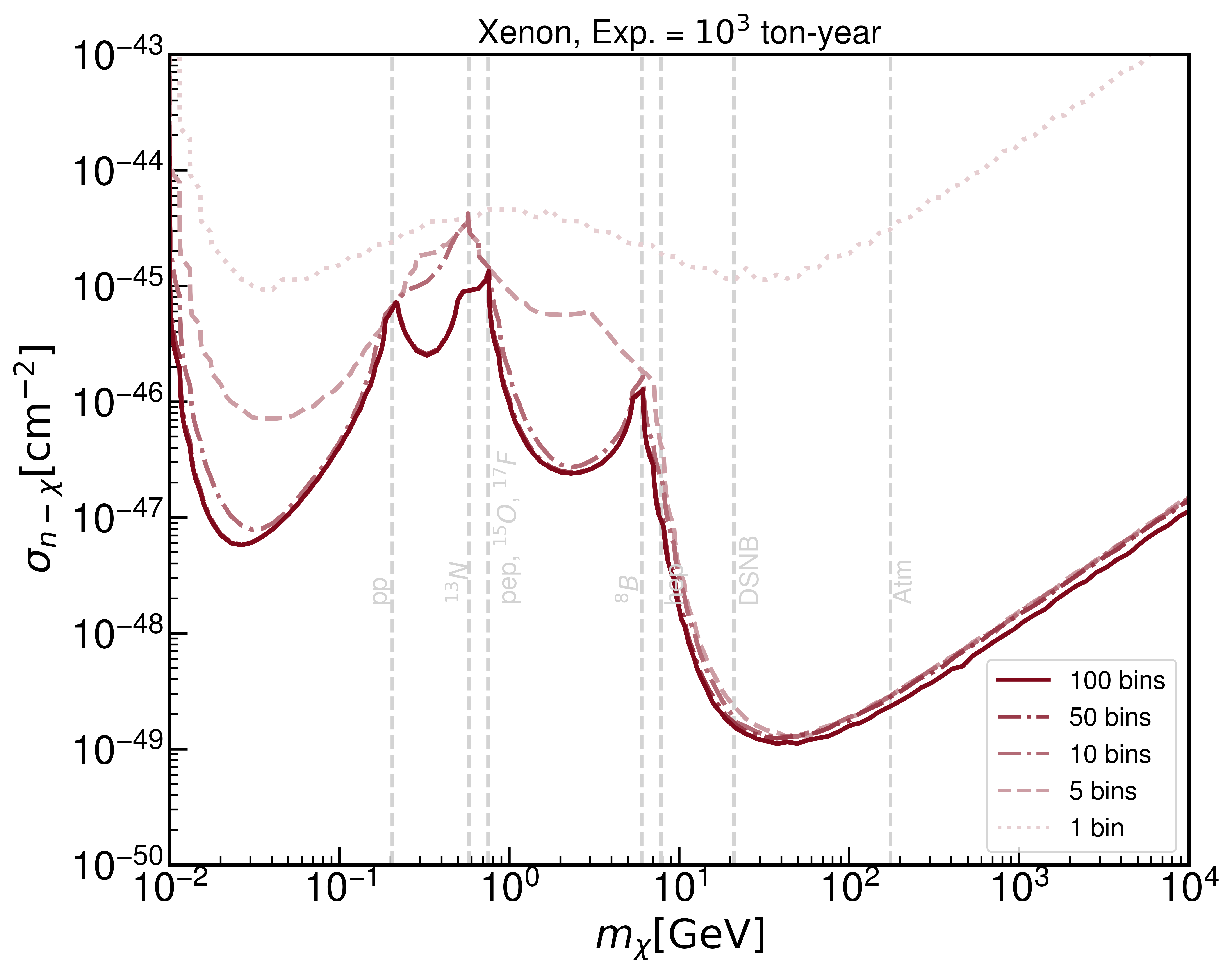}
\caption{Modification of the neutrino fog in the WIMP-nucleon SI cross-section, $\sigma_{n-\chi}$, versus WIMP mass, $m_\chi$ with the number of bins used in the analysis for a xenon-based detector in an Isospin-conserving case ($f_{n}/f_{p}=1$). On the left one ton-year exposure and on the right 1 kton-year exposure.}
\label{Fig:binninb}
\end{figure}

Regarding the binning, we are considering that the DM experiments are capable of having a large number of bins, otherwise the resolution to DM signal would be severely diminished as illustrated in Fig.~(\ref{Fig:binninb}) where we illustrate how an experiment with a binning smaller to 50 bins will loose resolution over the broad DM mass range: from $10^{-2}$ to $10^{4}$~GeV.
   
After this discussions on the needed elements to compute the neutrino fog, we can finally turn our attention into our case of interest: how the IVDM models shift the neutrino fog.
% ----------- IVDM.tex -----------

\section{\label{sec:IVDM}  Isospin Violating Dark Matter Models}

After describing the indispensable technical details for our computations, we focus on two IVDM realisations that illustrate how a particular DM model may cause a strong influence on the neutrino fog prediction.
 
Within the IVDM framework, DM couples differently to protons ($f_p$) and neutrons ($f_n$), departing from the assumption in standard WIMP scenarios. Allowing for unequal proton and neutron couplings introduces an Isospin asymmetry that can lead to destructive interference.  As a result, deviations from $f_p = f_n$ can substantially suppress the expected scattering rate for specific detector materials. Previous studies have examined the so-called xenophobic case, $f_n/f_p \approx -0.7$, in which destructive interference weakens xenon-based constraints and helps reconcile null xenon results with the DAMA/LIBRA and CoGeNT observations~\cite{Feng_2011}, which are in clear tension with more recent data from COSINE-100~\cite{COSINE-100:2024nfa}.

There are several DM models in which Isospin-violation arises naturally, for instance, some supersymmetric realisations~\cite{ZhaofengKang_2011,Lozano:2015vlv,Feng:2013vaa},  $Z'$-DM portal models~\cite{Frandsen:2011cg}, Two Higgs Doublet DM portal~\cite{Drozd:2015gda,Chang:2017gla}, among others~\cite{He:2016mls,He:2011gc}.   We have focused on two specific realisations of IVDM: the Scotogenic and the effective $Z$ portal DM model. The first is a popular model in which the DM phenomenology and the origin of the small neutrino masses are related. Moreover, the Scotogenic scheme involves a family of models that are of the most relevance for DM and neutrino phenomenology. For this reason, several extensions of this model  have been proposed recently, remarking its importance nowadays. The latter, although it is a simpler realisation, is ruled out by data; still, the origin of the Isospin violation is straightforward, thus this model gives a clear example to show the effects of this asymmetry in the neutrino fog.

\subsection{Scotogenic model}

\begin{figure}
\centering
\includegraphics[width=0.35\textwidth]{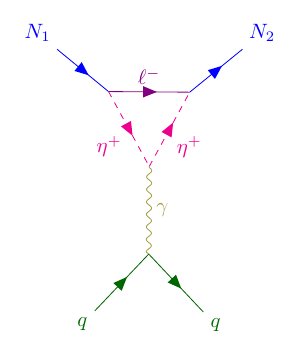}
\includegraphics[width=0.35\textwidth]{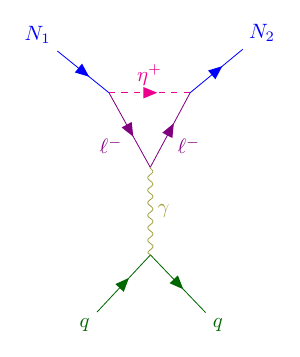}
\caption{Feynman diagrams for the SI inelastic scattering process of DM ($N_1$) off a nucleus at the lowest loop level in the Scotogenic model. This process is mediated via a photon and has been calculated in~\cite{Schmidt:2012yg}. }\label{fig:FeyScoto}
\end{figure}

The Scotogenic model~\cite{Ma:2006km, Tao:1996vb} is a minimal extension of the SM. In addition to the SM particle content, an extra inert scalar doublet with hyper-charge $Y = 1/2$, $\eta = (\eta^\pm, \eta_0 + i \eta_A )^T $, and $(i=1,2,3)$ generation of RH neutrinos $N_i$ are added. These RH neutrinos have a Majorana mass term $M_i$. The new fields transform under a global symmetry $\mathbb{Z}_2$, while the SM model particles transform trivially.  The relevant part of the Lagrangian is given by 
\begin{equation}
\mathcal{L} \subset - \frac{M_i}{2} \bar{N^c_i} P_R N_i + h_{\alpha i} \bar{\ell}_\alpha \eta^\dagger P_R N_i - V + \text{h.c.} ,
\end{equation}
where $P_R = \frac{1}{2} \left(1 +  \gamma_5 \right)$ is the right projector, $h_{\alpha i}$ are the Yukawa couplings and $V$ the scalar potential.

The Scotogenic model has scalar or fermionic DM candidates depending on the mass spectra. For scalar candidates, the DM phenomenology is similar to that of the IDM~\cite{LopezHonorez:2006gr,LopezHonorez:2010tb}, where the $\eta_0$ and $\eta_A$ couple in the same way to neutrons and protons at tree level; therefore, no Isospin violating couplings exist. On the other hand, fermionic DM, $N_i$, has no interaction with quarks at tree level; the model is leptophilic.  However, at the one-loop level, there is a coupling between the quarks and the lightest RH neutrino $N_1$ mediated by the photon, the $Z$ boson, and the Higgs~\cite{Ibarra:2016dlb}. From these interactions, the ones mediated by the photon and the $Z$ boson are SD, as they take the form as effective operator: $\bar{N}_1 \gamma^\mu \gamma_5 N_1 \, \bar{q} \gamma_\mu \gamma_5 q$. Lastly,  the interaction via the Higgs, although it generates a SI interaction, couples to the u and d-quarks via a Yukawa interaction; therefore, the Isospin violation is proportional to the mass difference, therefore becomes negligible.

However, as pointed out in~\cite{Schmidt:2012yg}, it is possible to realise a SI interaction that violates Isospin at the one-loop level in the case where the RH neutrinos $N_1$ and $N_2$ are almost degenerate. This consists of the inelastic scattering of $N_1$ into $N_2$ off the nucleus through the photon; the Feynman diagrams for this process are shown in Fig.~\ref{fig:FeyScoto}.
In this case, the SI inelastic scattering cross section occurs through the effective Lagrangian 
\begin{equation}
 \mathcal{L}_\text{eff}  = i a_{12} \bar{N_2} \gamma^\mu N_1 \partial^\nu F_{\nu \mu} 
 + i \left( \frac{\mu_{12}}{2} \right) \bar{N}_2 \sigma^{\mu \nu} N_1 F_{\mu \nu} 
 + i c_{12} \bar{N}_2 \gamma^\mu  N_1 A_\mu
\end{equation}
 where the functions are 
 \begin{equation}
 a_{12} = - \sum_a \frac{ \text{Im} ( h_{\alpha 2}^* h_{\alpha 1}) e }{ 2 (4 \pi) M^2_\eta} I_a \left( \frac{M_1^2}{M_\eta^2}, \frac{m_\alpha^2}{M_\eta^2} \right), 
 \end{equation}
  \begin{equation}
 \mu_{12} = - \sum_a \frac{ \text{Im} ( h_{\alpha 2}^* h_{\alpha 1}) e }{ 2 (4 \pi) M^2_\eta} 2 M_1 I_m \left( \frac{M_1^2}{M_\eta^2}, \frac{m_\alpha^2}{M_\eta^2} \right), 
 \end{equation}
  \begin{equation}
 c_{12} = \sum_a \frac{ \text{Im} ( h_{\alpha 2}^* h_{\alpha 1}) e }{ 2 (4 \pi) M^2_\eta} q^2 I_c \left( \frac{M_1^2}{M_\eta^2}, \frac{m_\alpha^2}{M_\eta^2} \right), 
 \end{equation}
 with $q$ the momentum transfer and $I_{a, m, c}$ the loop functions defined in~\cite{Schmidt:2012yg}.
 
Here, $f_n = 0$, because, in this case, the DM candidate $N_1$ couples only to $u$ and $d$-type quarks through the photon, then the asymmetry is $f_n/f_p  = 0$. Other choices in the parameter regions could give raise to different leading contributions to the SI cross-section. However, we have used this, somehow fine-tunned realisation of the model, to exemplify the limit case where DM couples only to protons in the nucleus. In contrast, in the model in the next subsection, the effective $Z$ portal DM, the contribution to the SI cross-section is maily due to a DM-neutron coupling.
 
\subsection{Effective $Z$ portal DM}

\begin{figure}
\centering
 \includegraphics[width=0.4\textwidth]{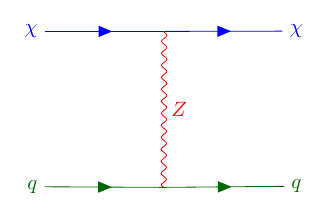}
\caption{Feynman diagram for the DM ($\chi$) -quark ($q$) elastic scattering in the effective $Z$-portal DM model.}\label{fig:FeyZ}
\end{figure}

The effective $Z$ portal DM model is a simple extension of the SM in which the $Z$ boson serves as the mediator between the dark and visible sectors. This framework is considered a minimal setup, as the $Z$ boson and the Higgs boson are the only particles within the SM that can serve as mediators between these sectors. In this model, a DM Dirac fermion particle, $\chi$, couples to the $Z$ boson, which then interacts with the fermions in the SM, $f$.
The general Lagrangian that describes these interactions is given by~\cite{Arcadi_2015}:
\begin{equation}
\mathcal{L} = \frac{g}{4 \cos \theta_W} \left( \bar{\chi} \, \gamma^{\mu} \left(V_\chi - A_\chi \gamma^5 \right) \chi \, Z_\mu + \bar{f} \, \gamma^{\mu} \left( V_f - A_f \gamma^5 \right) f \,Z_\mu \right).
\end{equation}
Here, $g$ is the weak coupling constant, $\theta_W$ is the weak mixing angle, and $V_{\chi, f}$ and $A_{\chi, f}$ represent the vectorial and axial couplings for the DM $(\chi)$ and SM fermions ($f$), respectively. 

In the case of a pure SI interaction, we take the axial part of the interaction as negligible, $A_\chi$, then the DM-nucleon interaction is through the $t$-channel exchange of a $Z$ boson, as shown in the Feynman diagram in Fig.~\ref{fig:FeyZ}.  Then, we are able to calculate the ratio of asymmetry, as the ratio of the neutron-DM to proton-DM couplings at tree level, giving
\begin{equation}
\frac{f_n}{f_p} = \frac{2 V_d + V_u}{V_d + 2 V_u} = - \frac{1}{1-4 \sin^2 \theta_W} \approx - 22.18,
\end{equation}
where the vector quark-Z couplings are given by $V_u = \frac{1}{2} - \left(\frac{4}{3} \right) \sin^2 \theta_W $ and $ V_d = - \frac{1}{2} + \left(\frac{2}{3} \right) \sin^2 \theta_W$ and the weak mixing angle (in the $\mathrm{\bar{MS}}$ Scheme at low energies) is taken as $ \sin^2 \theta_W = 0.23873 \pm 0.00005$~\cite{ParticleDataGroup:2024cfk, Erler:2017knj}.

The effective $Z$-portal DM model provides a direct example of Isospin violation in which the coupling ratio is fixed and determined by the weak mixing angle. Since our goal is only to consider the effect of Isospin violation on the neutrino fog, an effective $Z$-portal model allows us to demonstrate the impact of a large Isospin violation without necessarily affecting the standard neutrino background. The effect of adding a $Z'$ mediator to the neutrino fog has been studied in~\cite{lozano2025}.

\subsection{Neutrino Fog in IVDM models}

\begin{figure}[htb]
    \centering
    \includegraphics[width=0.7\textwidth]{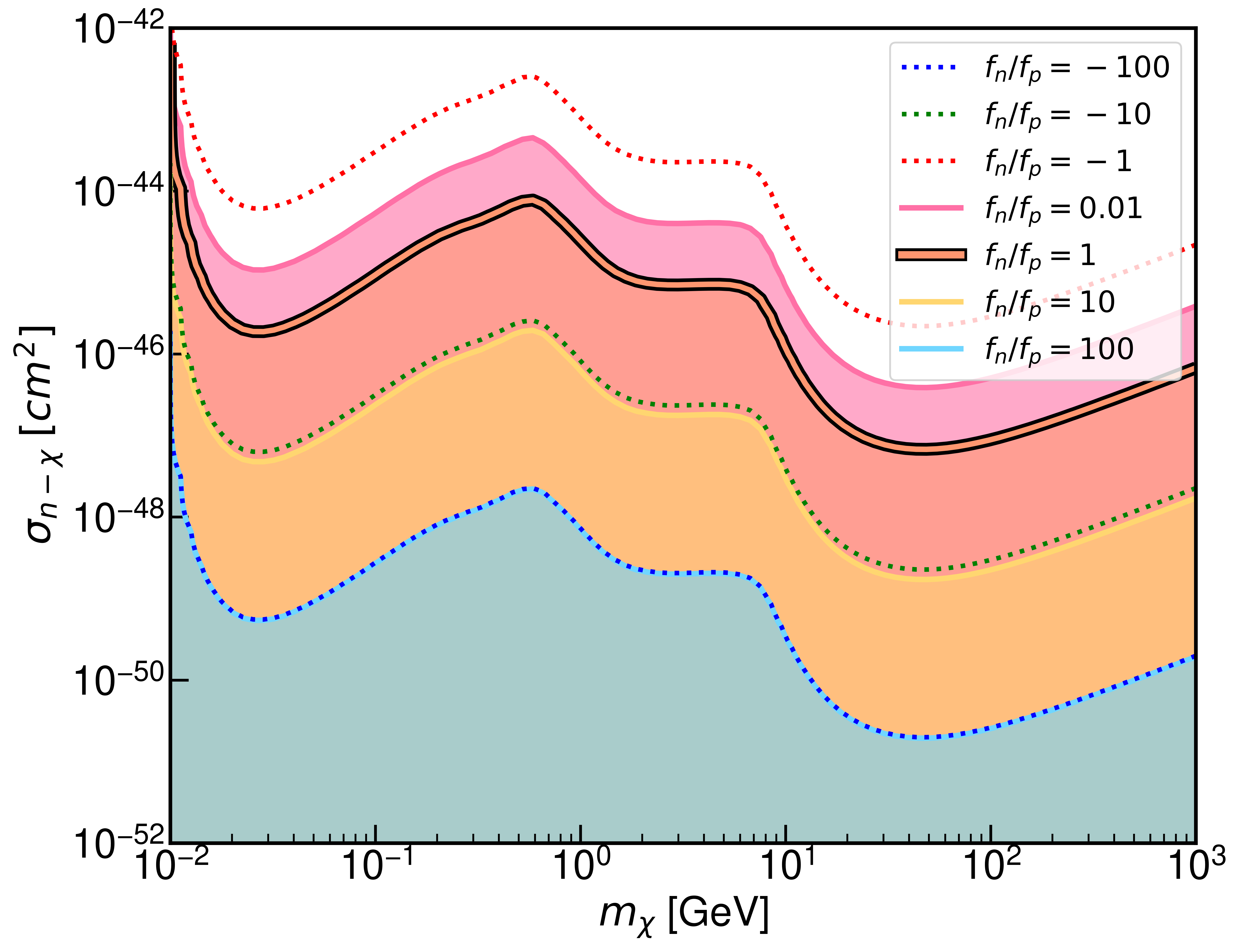}
    \caption{The discovery limit for a xenon-based detector plotted in the WIMP mass ($m_\chi$) versus WIMP-nucleon SI cross-section ($\sigma_{n-\chi}$) plane for different values of the ratio $f_n/f_p$. The standard Isospin-conserving scenario, $f_n/f_p=1$, is indicated by the thick black line within the peach-coloured region. }
    \label{Fig:Ratios}
\end{figure}

Motivated by the above discussed IVDM models, we can compute the expected neutrino fog region in terms of the ratio $f_n/f_p$, to have an expectation for a whole family of IVDM models, including the two models described above. To perform such a computation, we start from Eq.~\ref{eq:sigma}, where the DM event rate is proportional to the square of the sum of the contributions from protons and neutrons in the nucleus. This expression can be rewritten as
 \begin{equation} \label{Eq:Ratio}
 R_{DM} \propto \left[ Z + \left(\frac{f_n}{f_p}\right)(A-Z)\right]^2 .
 \end{equation}
The value and sign of $f_n/f_p$ can change the balance of the contribution of each of the nucleons, or determine whether these contributions add up or cancel each other out, as shown in Fig.~\ref{Fig:Ratios}. The sign will determine the type of interference. When the sign is positive, the contributions of protons and neutrons add up to each other. This results in constructive interference, increasing the DM event rate, improving the sensitivity of potential detectors and pushing the discovery limit to smaller cross-sections. 
On the other hand, a negative sign generates destructive interference. In this case, the contributions of protons and neutrons cancel each other, suppressing DM events. This reduces the detector sensitivity and shifts the discovery limit to larger cross-sections.\\
For those elements with one isotope that dominates the abundance, there is a specific negative ratio value that generates perfect cancellation. In the case of elements with multiple isotopes, complete destructive interference cannot be achieved for all isotopes simultaneously, so only partial cancellation can be generated. For xenon, the value of the ratio between the neutron and proton coupling strengths that causes maximum suppression is $f_n/f_p \approx -0.7$~\cite{Feng_2013}. The Fig.~\ref{Fig:Ratios} shows the case for $f_n/f_p = -1$, and although this value does not yield maximum cancellation for xenon, it is the closest to $-0.7$ among the scenarios presented, leading to a higher neutrino fog.\\

The value of the ratio $f_n/f_p$ determines which part of the nucleus dominates the interaction. A large ratio would imply neutron dominance, which can be realised in several IVDM models without a fine tuning. In this case, the sign of the ratio is not particularly relevant, since this term is squared. As ilustration we show that both ratios, $-100$ and $100$, produce a large number of events, making the detector very sensitive. On the contrary, if the magnitude of the ratio is small, both neutrons and protons contribute similarly. They can combine perfectly, as in the standard case for a ratio of $1$, or they can perfectly cancel each other out. 

\begin{figure}[t]
\includegraphics[width=.49\linewidth]{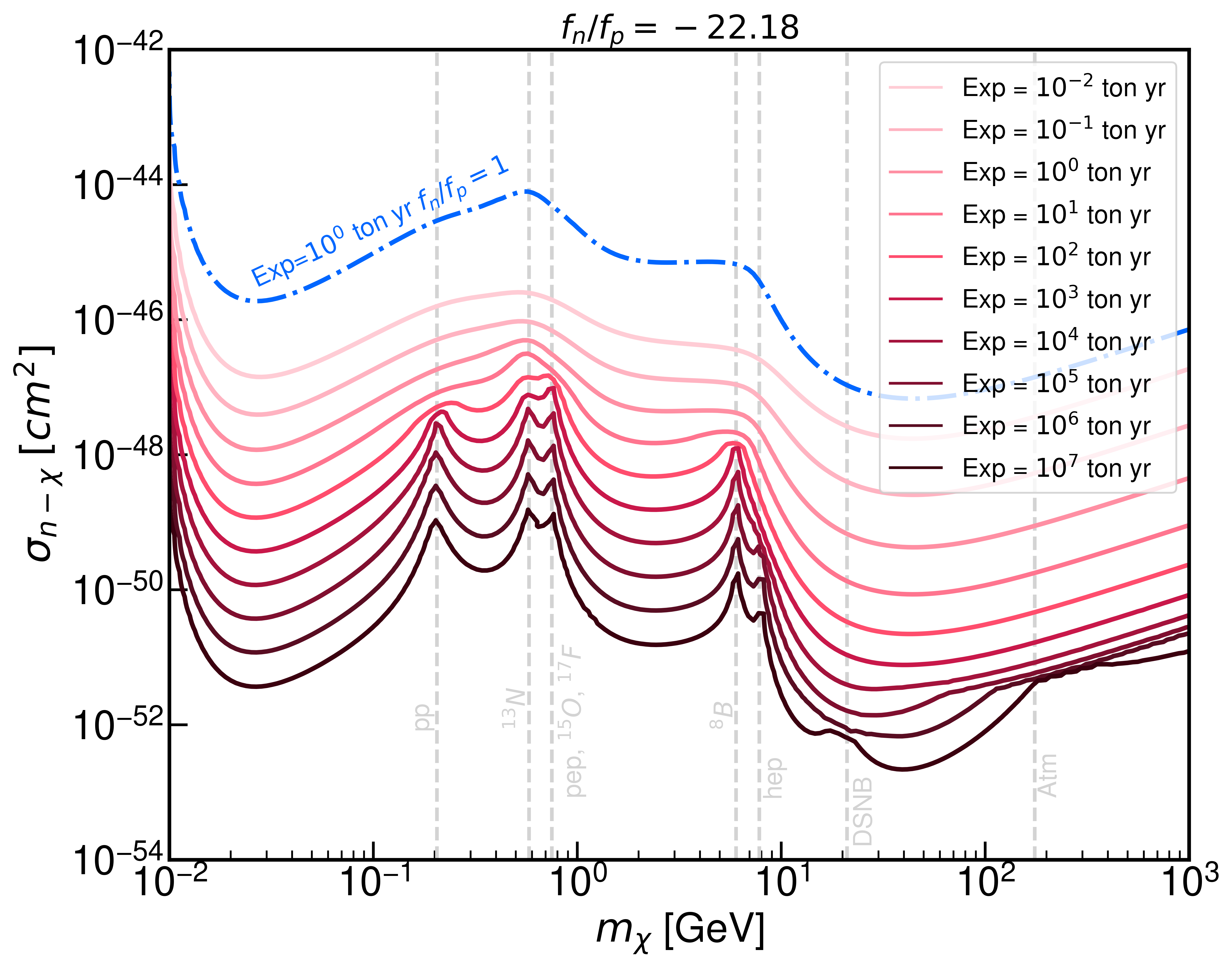}\hfill
\includegraphics[width=.49\linewidth]{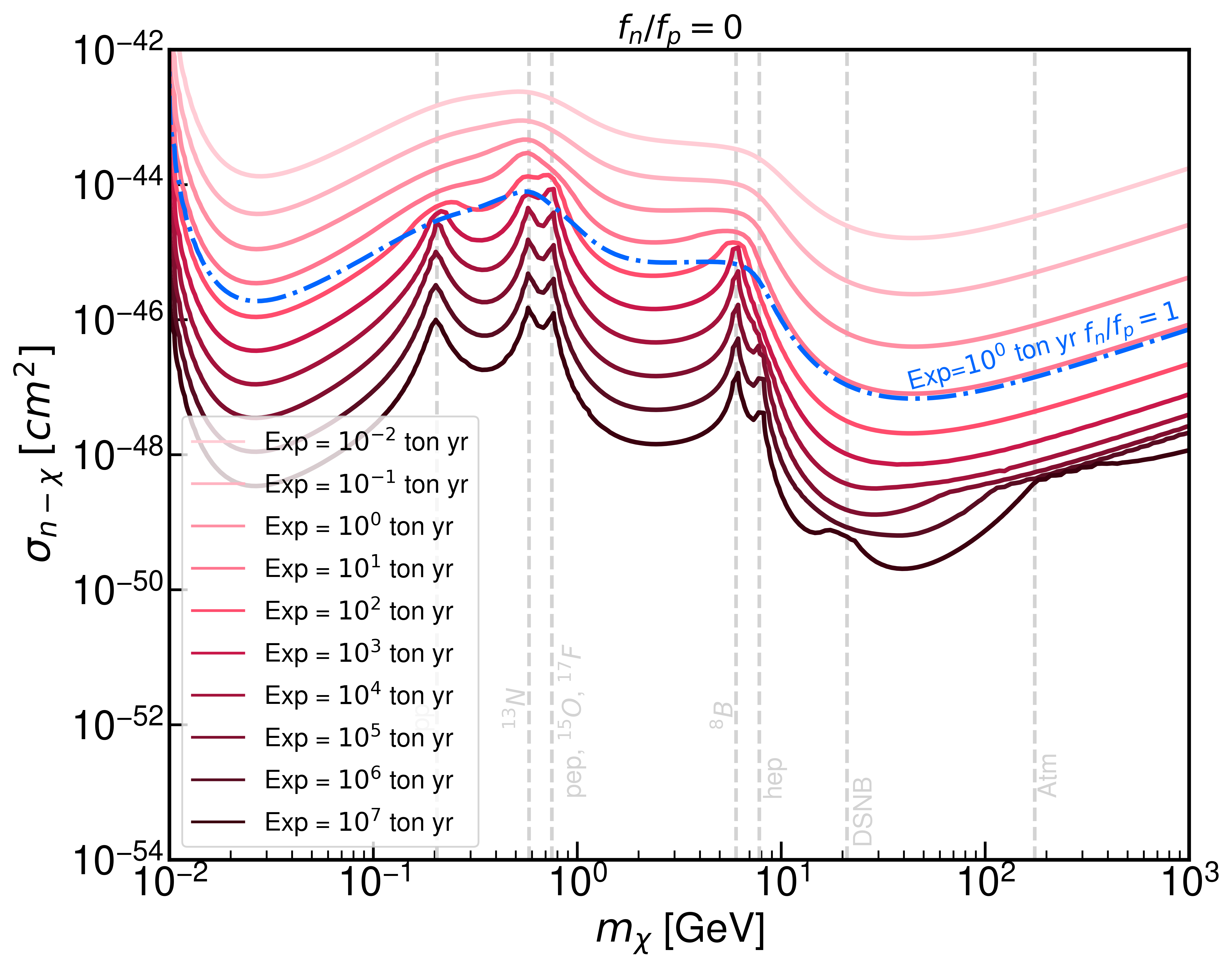}
\caption{Evolution of the neutrino fog in the WIMP-nucleon SI cross-section, $\sigma_{n-\chi}$, versus WIMP mass, $m_\chi$, plane for a xenon-based detector. The solid curves represent discovery limits for experimental exposures ranging from $10^{-2}$ ton-year (light pink) to $10^7$ ton-year (dark maroon). The panels compare two Isospin-violating scenarios: the $Z$-portal DM model with $f_n / f_p=-22.18$ (left) and the Scotogenic model realization with $f_n/ f_p=0$ (right). For reference, the blue dash-dotted line indicates the projected limit for the standard Isospin-conserving case ($f_{n}/f_{p}=1$) at one ton-year exposure.}
\label{Fig:Exp}
\end{figure}

To visualise the impact of the two scenarios presented above, Fig.~\ref{Fig:Exp} shows the neutrino fog for a xenon detector and evaluates the effect of changes in the ratio predicted by each model. The solid curves indicate the discovery limits as experimental exposure increases. The left panel corresponds to the effective $Z$-portal DM model ($f_n/f_p = -22.18$), in which the large negative ratio enhances the DM event rate due to the neutron rich composition of the xenon nucleus, enabling the experiment to test significantly lower cross-sections before the experiment reaches sensitivity to neutrinos, shifting the neutrino fog to smaller cross-sections than the standard case of Isospin conservation ($f_n/f_p=1$)   presented as a reference. For instance, at a fixed mass of $100$ GeV and an exposure of $1$ ton-year, the projected discovery limit in the $Z$-portal model improves by a factor of $159$ relative to the Isospin-conserving case. On the other hand, the right panel shows the Scotogenic model realization with $f_n/f_p = 0$, which couples only to protons at one-loop level; this decoupling from neutrons can suppress the DM event rate, thereby shifting the neutrino fog to larger cross-sections: at the same fixed mass of $100$ GeV and one ton-year exposure, the projected discovery limit is lowered by a factor of $0.17$ compared to the Isospin conserving case.

\begin{figure}
    \centering
    \includegraphics[width=0.7\textwidth]{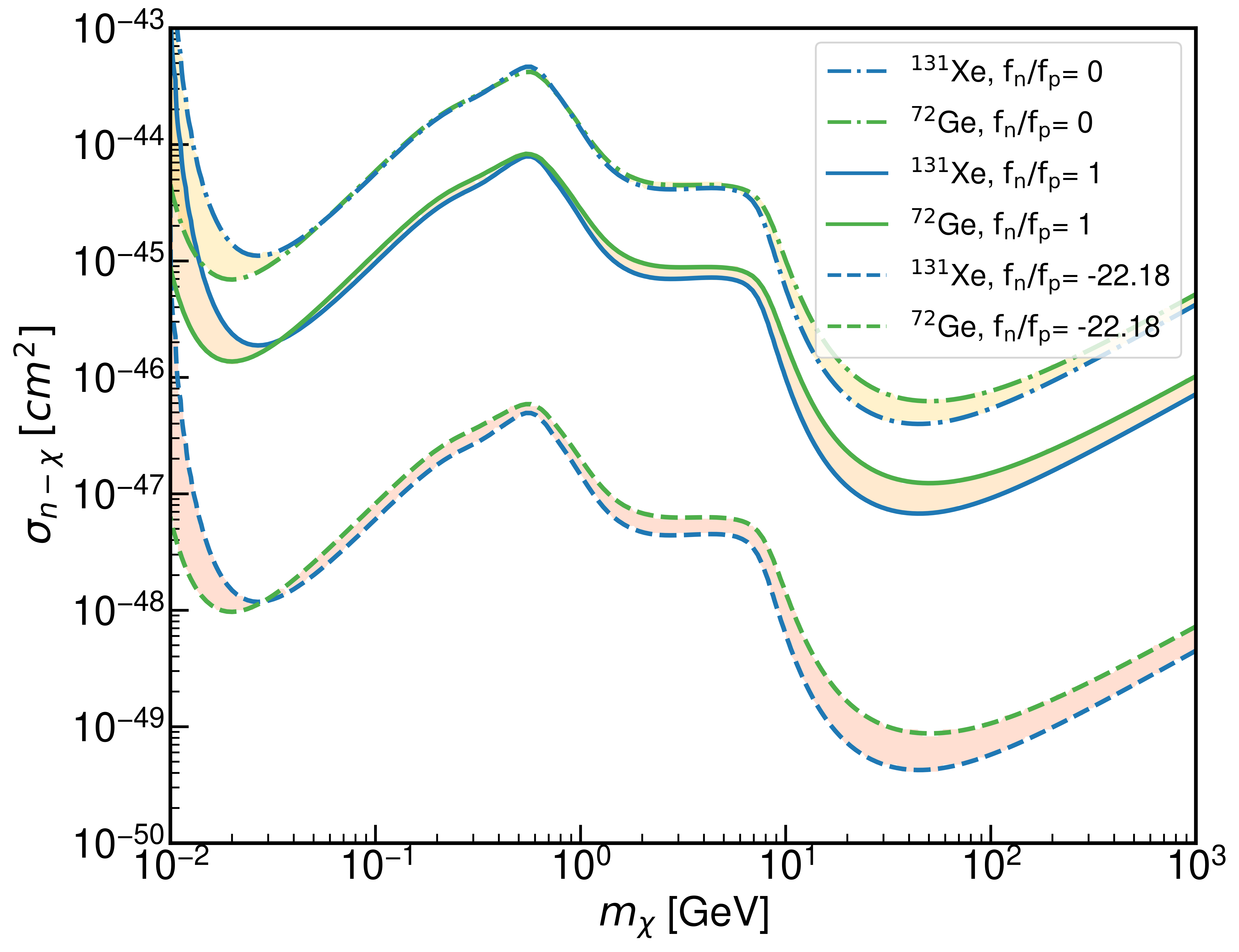}
    \caption{Comparison of the neutrino fog for two target materials in the WIMP-nucleon SI cross-section ($\sigma_{n-\chi}$)  versus WIMP mass ($m_\chi$) plane: $^{131}$Xe (blue lines) and $^{72}$Ge (green lines). The different line styles correspond to three different $f_n/f_p$ ratios: $f_n/f_p = 0$ (dash-dotted lines), the standard Isospin-conserving case $f_n/f_p=1$ (solid lines), and $f_n/f_p=-22.18$ (dashed lines). The plot shows that the relative sensitivity between the two target materials depends strongly on the specific $f_n/f_p$ ratio.}
    \label{Fig:Material}
\end{figure}

Finally, Fig.~\ref{Fig:Material} highlights the important role of the target material composition in establishing the neutrino fog in the different Isospin-violation scenarios we have presented. This plot shows a comparison between a $^{131}$Xe target and a $^{72}$Ge target. Xenon generally produces a neutrino fog prediction at smaller cross-sections than germanium, this result is to be expected because the DM event rate scales according to Eq.~\ref{Eq:Ratio}. In the standard Isospin-conserving scenario the cross-section scales with the square of the number of nucleons ($A^2$), wich naturally favours the heavier xenon nucleus. 
In the presented Scotogenic model realization with $f_n/f_p=0$, the discovery limits for both targets improve relative to the standard scenario, and the difference between the discovery limits of xenon and germanium decreases, resulting in comparative rates of $R_{DM}\propto 3000$ for Xe and $R_{DM}\propto 1000$ for Ge. Meanwhile, the $Z$-portal DM model exhibits an effective coupling that strongly couples to neutrons. Since xenon contains a higher number of neutrons $(A-Z)$, it benefits from this scaling, resulting in a higher event rate ($R_{DM}\propto 2.73 \times 10^6$ for Xe compared to $R_{DM}\propto 7.31 \times 10^5$ for Ge). As a result, the neutrino fog for xenon \textbf{shifts to much smaller cross-sections compared to} germanium, creating a wider gap between the sensitivity of these materials.

Therefore, these results highlight the importance of complementary targets in future DD experiments. As shown, in scenarios such as the $Z$-portal DM model, materials with a higher neutron content, such as xenon, exhibit a neutrino fog at lower cross sections than that for germanium. This material-dependent response implies that rather than relying on a single material as a detector, an approach combining isotopic compositions and different materials may be crucial for fully exploring the parameter space.

% ----------- Conclusions.tex -----------

\section{\label{sec:Conclusions}  Conclusions}

In this work, we have studied the impact of IVDM models, such as the Scotogenic and effective $Z$-portal DM models, on the projected sensitivity of DM DD experiments, focusing on variations in the coupling ratio $f_n/f_p$. The results presented  illustrate the variable character of the neutrino fog and show that it shifts significantly depending on the underlying particle physics.
For the concrete models we use as examples, we have shown that the effective $Z$-portal DM model, which has a strong coupling to neutrons, produces a downward shift in the neutrino fog, thereby improving experimental sensitivity. The Scotogenic model, on the other hand, being a model that decouples from neutrons, results in a higher neutrino fog, requiring high exposure values in order to be tested.\\

Contrary to Isospin-violating scenarios considered to reconcile conflicting experimental data (e.g., the xenophobic case), the coupling ratios studied here arise naturally from the proposed DM models, demonstrating that significant shifts in the neutrino fog can be direct predictions of physically motivated extensions of the SM.\\

Finally, we have shown the impact of the IVDM models on the neutrino fog for different target materials. We have compared the xenon and germanium cases, showing that differences in sensitivity between the materials depend on the model under study. These results then highlight that a multi-target approach will be essential not only for discovery but for determining the precise coupling properties of the DM.

\begin{acknowledgments}
We would like to thank Luis Jorge Flores for his time and valuable discussion. OGM and JML would like to thank Sistema Nacional de Investigadoras e Investigadores (SNII). JML is founded by the Secihti Estancias posdoctorales por México 2023 grant. 
\end{acknowledgments}

%\appendix
%\section{Appendixes}

\bibliographystyle{apsrev} 
\bibliography{bibliography}

@PREAMBLE{
 "\providecommand{\noopsort}[1]{}" 
 # "\providecommand{\singleletter}[1]{#1}%" 
}

@article{osti_4701226,
    author = "Lindhard, Nielsen, Scharff and Thomsen, P.~V.",
    title = "{Integral Equations Governing Radiation Effects (Notes on Atomic Collisions, III)}",
    journal = "Kgl. Danske Videnskabernes Selskab, Matematisk-Fysiske Meddelelser",
    volume = "22",
    pages = "10",
    year = "1963"
}

@article{Catena:2015uua,
    author = "Catena, Riccardo and Gondolo, Paolo",
    title = "{Global limits and interference patterns in dark matter direct detection}",
    eprint = "1504.06554",
    archivePrefix = "arXiv",
    primaryClass = "hep-ph",
    doi = "10.1088/1475-7516/2015/08/022",
    journal = "JCAP",
    volume = "08",
    pages = "022",
    year = "2015"
}

@article{Brenner:2022qku,
    author = "Brenner, Anja and Herrera, Gonzalo and Ibarra, Alejandro and Kang, Sunghyun and Scopel, Stefano and Tomar, Gaurav",
    title = "{Complementarity of experiments in probing the non-relativistic effective theory of dark matter-nucleon interactions}",
    eprint = "2203.04210",
    archivePrefix = "arXiv",
    primaryClass = "hep-ph",
    doi = "10.1088/1475-7516/2022/06/026",
    journal = "JCAP",
    volume = "06",
    number = "06",
    pages = "026",
    year = "2022"
}

@article{AvisKozar:2023iyb,
    author = "Avis Kozar, Neal P. and Scott, Pat and Vincent, Aaron C.",
    title = "{A global fit of non-relativistic effective dark matter operators including solar neutrinos}",
    eprint = "2310.15392",
    archivePrefix = "arXiv",
    primaryClass = "hep-ph",
    doi = "10.1088/1475-7516/2025/02/007",
    journal = "JCAP",
    volume = "02",
    pages = "007",
    year = "2025"
}

@article{Sorensen:2014sla,
    author = "Sorensen, Peter",
    title = "{Atomic limits in the search for galactic dark matter}",
    eprint = "1412.3028",
    archivePrefix = "arXiv",
    primaryClass = "astro-ph.IM",
    doi = "10.1103/PhysRevD.91.083509",
    journal = "Phys. Rev. D",
    volume = "91",
    number = "8",
    pages = "083509",
    year = "2015"
}

@article{Chang:2017gla,
    author = "Chang, Chia-Feng and He, Xiao-Gang and Tandean, Jusak",
    title = "{Two-Higgs-Doublet-Portal Dark-Matter Models in Light of Direct Search and LHC Data}",
    eprint = "1702.02924",
    archivePrefix = "arXiv",
    primaryClass = "hep-ph",
    reportNumber = "NCTS-PH-1707",
    doi = "10.1007/JHEP04(2017)107",
    journal = "JHEP",
    volume = "04",
    pages = "107",
    year = "2017"
}

@article{He:2016mls,
    author = "He, Xiao-Gang and Tandean, Jusak",
    title = "{New LUX and PandaX-II Results Illuminating the Simplest Higgs-Portal Dark Matter Models}",
    eprint = "1609.03551",
    archivePrefix = "arXiv",
    primaryClass = "hep-ph",
    doi = "10.1007/JHEP12(2016)074",
    journal = "JHEP",
    volume = "12",
    pages = "074",
    year = "2016"
}

@article{He:2011gc,
    author = "He, Xiao-Gang and Ren, Bo and Tandean, Jusak",
    title = "{Hints of Standard Model Higgs Boson at the LHC and Light Dark Matter Searches}",
    eprint = "1112.6364",
    archivePrefix = "arXiv",
    primaryClass = "hep-ph",
    doi = "10.1103/PhysRevD.85.093019",
    journal = "Phys. Rev. D",
    volume = "85",
    pages = "093019",
    year = "2012"
}

@article{Carew:2023qrj,
    author = "Carew, Ben and Caddell, Ashlee R. and Maity, Tarak Nath and O'Hare, Ciaran A. J.",
    title = "{Neutrino fog for dark matter-electron scattering experiments}",
    eprint = "2312.04303",
    archivePrefix = "arXiv",
    primaryClass = "hep-ph",
    doi = "10.1103/PhysRevD.109.083016",
    journal = "Phys. Rev. D",
    volume = "109",
    number = "8",
    pages = "083016",
    year = "2024"
}

@article{Essig:2018tss,
    author = "Essig, Rouven and Sholapurkar, Mukul and Yu, Tien-Tien",
    title = "{Solar Neutrinos as a Signal and Background in Direct-Detection  Experiments Searching for Sub-GeV Dark Matter With Electron Recoils}",
    eprint = "1801.10159",
    archivePrefix = "arXiv",
    primaryClass = "hep-ph",
    reportNumber = "YITB-SB-17-36, YITP-SB-17-36, CERN-TH-2017-194",
    doi = "10.1103/PhysRevD.97.095029",
    journal = "Phys. Rev. D",
    volume = "97",
    number = "9",
    pages = "095029",
    year = "2018"
}

@article{Herrera:2023xun,
    author = "Herrera, Gonzalo",
    title = "{A neutrino floor for the Migdal effect}",
    eprint = "2311.17719",
    archivePrefix = "arXiv",
    primaryClass = "hep-ph",
    doi = "10.1007/JHEP05(2024)288",
    journal = "JHEP",
    volume = "05",
    pages = "288",
    year = "2024"
}

@article{Maity:2024hzb,
    author = "Maity, Tarak Nath",
    title = "{Neutrinos as background and signal in searches using the Migdal effect}",
    eprint = "2412.17649",
    archivePrefix = "arXiv",
    primaryClass = "hep-ph",
    doi = "10.1103/h3th-6wsr",
    journal = "Phys. Rev. D",
    volume = "111",
    number = "12",
    pages = "123020",
    year = "2025"
}

@article{ZhaofengKang_2011,
doi = {10.1088/1475-7516/2011/01/028},
url = {https://doi.org/10.1088/1475-7516/2011/01/028},
year = {2011},
month = {jan},
publisher = {},
volume = {2011},
number = {01},
pages = {028},
author = {Zhaofeng Kang and Tianjun Li and Tao Liu and Chunli Tong and Jin Min Yang},
title = {Light dark matter from the U(1)X sector in the NMSSM with gauge mediation},
journal = {Journal of Cosmology and Astroparticle Physics},
}

@article{Drozd:2015gda,
    author = "Drozd, Aleksandra and Grzadkowski, Bohdan and Gunion, John F. and Jiang, Yun",
    title = "{Isospin-violating dark-matter-nucleon scattering via two-Higgs-doublet-model portals}",
    eprint = "1510.07053",
    archivePrefix = "arXiv",
    primaryClass = "hep-ph",
    doi = "10.1088/1475-7516/2016/10/040",
    journal = "JCAP",
    volume = "10",
    pages = "040",
    year = "2016"
}

@article{Frandsen:2011cg,
    author = "Frandsen, Mads T. and Kahlhoefer, Felix and Sarkar, Subir and Schmidt-Hoberg, Kai",
    title = "{Direct detection of dark matter in models with a light Z'}",
    eprint = "1107.2118",
    archivePrefix = "arXiv",
    primaryClass = "hep-ph",
    reportNumber = "OUTP-11-44",
    doi = "10.1007/JHEP09(2011)128",
    journal = "JHEP",
    volume = "09",
    pages = "128",
    year = "2011"
}

@inproceedings{Feng:2013vaa,
    author = "Feng, Jonathan L. and Kumar, Jason and Marfatia, Danny and Sanford, David",
    title = "{Isospin-Violating Dark Matter Benchmarks for Snowmass 2013}",
    booktitle = "{Snowmass 2013}: {Snowmass on the Mississippi}",
    eprint = "1307.1758",
    archivePrefix = "arXiv",
    primaryClass = "hep-ph",
    month = "7",
    year = "2013"
}

@article{Lozano:2015vlv,
    author = "Lozano, V{\'\i}ctor Mart{\'\i}n and Peir{\'o}, Miguel and Soler, Pablo",
    title = {{Isospin violating dark matter in St{\"u}ckelberg portal scenarios}},
    eprint = "1503.01780",
    archivePrefix = "arXiv",
    primaryClass = "hep-ph",
    reportNumber = "FTUAM-15-5, IFT-UAM-CSIC-15-018, MAD-TH-15-03",
    doi = "10.1007/JHEP04(2015)175",
    journal = "JHEP",
    volume = "04",
    pages = "175",
    year = "2015"
}

@article{LopezHonorez:2006gr,
    author = "Lopez Honorez, Laura and Nezri, Emmanuel and Oliver, Josep F. and Tytgat, Michel H. G.",
    title = "{The Inert Doublet Model: An Archetype for Dark Matter}",
    eprint = "hep-ph/0612275",
    archivePrefix = "arXiv",
    reportNumber = "ULB-TH-06-27",
    doi = "10.1088/1475-7516/2007/02/028",
    journal = "JCAP",
    volume = "02",
    pages = "028",
    year = "2007"
}

@article{LopezHonorez:2010tb,
    author = "Lopez Honorez, Laura and Yaguna, Carlos E.",
    title = "{A new viable region of the inert doublet model}",
    eprint = "1011.1411",
    archivePrefix = "arXiv",
    primaryClass = "hep-ph",
    reportNumber = "ULB-TH-10-37",
    doi = "10.1088/1475-7516/2011/01/002",
    journal = "JCAP",
    volume = "01",
    pages = "002",
    year = "2011"
}

@article{Lewin:1995rx,
    author = "Lewin, J. D. and Smith, P. F.",
    title = "{Review of mathematics, numerical factors, and corrections for dark matter experiments based on elastic nuclear recoil}",
    reportNumber = "RAL-TR-95-024",
    doi = "10.1016/S0927-6505(96)00047-3",
    journal = "Astropart. Phys.",
    volume = "6",
    pages = "87--112",
    year = "1996"
}

@article{AristizabalSierra:2021kht,
    author = "Aristizabal Sierra, D. and De Romeri, V. and Flores, L. J. and Papoulias, D. K.",
    title = "{Impact of COHERENT measurements, cross section uncertainties and new interactions on the neutrino floor}",
    eprint = "2109.03247",
    archivePrefix = "arXiv",
    primaryClass = "hep-ph",
    doi = "10.1088/1475-7516/2022/01/055",
    journal = "JCAP",
    volume = "01",
    number = "01",
    pages = "055",
    year = "2022"
}

@article{Billard,
  title = {Implication of neutrino backgrounds on the reach of next generation dark matter direct detection experiments},
  author = {Billard, J. and Figueroa-Feliciano, E. and Strigari, L.},
  journal = {Phys. Rev. D},
  volume = {89},
  issue = {2},
  pages = {023524},
  numpages = {15},
  year = {2014},
  month = {Jan},
  publisher = {American Physical Society},
  doi = {10.1103/PhysRevD.89.023524},
  url = {https://link.aps.org/doi/10.1103/PhysRevD.89.023524}
}

@article{Billard:2011zj,
    author = "Billard, J. and Mayet, F. and Santos, D.",
    title = "{Assessing the discovery potential of directional detection of Dark Matter}",
    eprint = "1110.6079",
    archivePrefix = "arXiv",
    primaryClass = "astro-ph.CO",
    doi = "10.1103/PhysRevD.85.035006",
    journal = "Phys. Rev. D",
    volume = "85",
    pages = "035006",
    year = "2012"
}

@article{Cowan:2010js,
    author = "Cowan, Glen and Cranmer, Kyle and Gross, Eilam and Vitells, Ofer",
    title = "{Asymptotic formulae for likelihood-based tests of new physics}",
    eprint = "1007.1727",
    archivePrefix = "arXiv",
    primaryClass = "physics.data-an",
    doi = "10.1140/epjc/s10052-011-1554-0",
    journal = "Eur. Phys. J. C",
    volume = "71",
    pages = "1554",
    year = "2011",
    note = "[Erratum: Eur.Phys.J.C 73, 2501 (2013)]"
}

@article{Savage,
  title = {Annual modulation of dark matter in the presence of streams},
  author = {Savage, Chris and Freese, Katherine and Gondolo, Paolo},
  journal = {Phys. Rev. D},
  volume = {74},
  issue = {4},
  pages = {043531},
  numpages = {16},
  year = {2006},
  month = {Aug},
  publisher = {American Physical Society},
  doi = {10.1103/PhysRevD.74.043531},
  url = {https://link.aps.org/doi/10.1103/PhysRevD.74.043531}
}

@article{Freese_2013,
   title={Colloquium: Annual modulation of dark matter},
   volume={85},
   ISSN={1539-0756},
   url={http://dx.doi.org/10.1103/RevModPhys.85.1561},
   DOI={10.1103/revmodphys.85.1561},
   number={4},
   journal={Reviews of Modern Physics},
   publisher={American Physical Society (APS)},
   author={Freese, Katherine and Lisanti, Mariangela and Savage, Christopher},
   year={2013},
   month=nov, pages={1561–1581} }

@article{Arcadi_2015,
doi = {10.1088/1475-7516/2015/03/018},
url = {https://dx.doi.org/10.1088/1475-7516/2015/03/018},
year = {2015},
month = {mar},
publisher = {},
volume = {2015},
number = {03},
pages = {018},
author = {Arcadi, Giorgio and Mambrini, Yann and Richard, Francois},
title = {Z-portal dark matter},
journal = {Journal of Cosmology and Astroparticle Physics}
}

@misc{lozano2025,
      title={Neutrino fog in the light dark sector: the role of isospin violation}, 
      author={Víctor Martín Lozano and Shankar Pramanik and Soumya Sadhukhan and Adrián Terrones},
      year={2025},
      eprint={2508.05787},
      archivePrefix={arXiv},
      primaryClass={hep-ph},
      url={https://arxiv.org/abs/2508.05787}, 
}

@article{ParticleDataGroup:2024cfk,
    author = "Navas, S. and others",
    collaboration = "Particle Data Group",
    title = "{Review of particle physics}",
    doi = "10.1103/PhysRevD.110.030001",
    journal = "Phys. Rev. D",
    volume = "110",
    number = "3",
    pages = "030001",
    year = "2024"
}

@article{Tao:1996vb,
    author = "Tao, Zhi-jian",
    title = "{Radiative seesaw mechanism at weak scale}",
    eprint = "hep-ph/9603309",
    archivePrefix = "arXiv",
    doi = "10.1103/PhysRevD.54.5693",
    journal = "Phys. Rev. D",
    volume = "54",
    pages = "5693--5697",
    year = "1996"
}

@article{Ma:2006km,
    author = "Ma, Ernest",
    title = "{Verifiable radiative seesaw mechanism of neutrino mass and dark matter}",
    eprint = "hep-ph/0601225",
    archivePrefix = "arXiv",
    reportNumber = "UCRHEP-T403",
    doi = "10.1103/PhysRevD.73.077301",
    journal = "Phys. Rev. D",
    volume = "73",
    pages = "077301",
    year = "2006"
}

@article{Ibarra:2016dlb,
    author = "Ibarra, Alejandro and Yaguna, Carlos E. and Zapata, Oscar",
    title = "{Direct Detection of Fermion Dark Matter in the Radiative Seesaw Model}",
    eprint = "1601.01163",
    archivePrefix = "arXiv",
    primaryClass = "hep-ph",
    reportNumber = "TUM-HEP-967-1, MS-TP-14-36, FLAVOUR(267104)-ERC-87",
    doi = "10.1103/PhysRevD.93.035012",
    journal = "Phys. Rev. D",
    volume = "93",
    number = "3",
    pages = "035012",
    year = "2016"
}

@article{Schmidt:2012yg,
    author = "Schmidt, Daniel and Schwetz, Thomas and Toma, Takashi",
    title = "{Direct Detection of Leptophilic Dark Matter in a Model with Radiative Neutrino Masses}",
    eprint = "1201.0906",
    archivePrefix = "arXiv",
    primaryClass = "hep-ph",
    reportNumber = "KANAZAWA-11-14",
    doi = "10.1103/PhysRevD.85.073009",
    journal = "Phys. Rev. D",
    volume = "85",
    pages = "073009",
    year = "2012"
}

@article{Erler:2017knj,
    author = "Erler, Jens and Ferro-Hern{\'a}ndez, Rodolfo",
    title = "{Weak Mixing Angle in the Thomson Limit}",
    eprint = "1712.09146",
    archivePrefix = "arXiv",
    primaryClass = "hep-ph",
    doi = "10.1007/JHEP03(2018)196",
    journal = "JHEP",
    volume = "03",
    pages = "196",
    year = "2018"
}

@article{Papoulias_2019,
   title={Recent Probes of Standard and Non-standard Neutrino Physics With Nuclei},
   volume={7},
   ISSN={2296-424X},
   url={http://dx.doi.org/10.3389/fphy.2019.00191},
   DOI={10.3389/fphy.2019.00191},
   journal={Frontiers in Physics},
   publisher={Frontiers Media SA},
   author={Papoulias, Dimitrios K. and Kosmas, Theocharis S. and Kuno, Yoshitaka},
   year={2019},
   month=nov }

@article{Klein_1999,
   title={Exclusive vector meson production in relativistic heavy ion collisions},
   volume={60},
   ISSN={1089-490X},
   url={http://dx.doi.org/10.1103/PhysRevC.60.014903},
   DOI={10.1103/physrevc.60.014903},
   number={1},
   journal={Physical Review C},
   publisher={American Physical Society (APS)},
   author={Klein, Spencer R. and Nystrand, Joakim},
   year={1999},
   month=jun }

@article{Baxter_2021,
   title={Recommended conventions for reporting results from direct dark matter searches},
   volume={81},
   ISSN={1434-6052},
   url={http://dx.doi.org/10.1140/epjc/s10052-021-09655-y},
   DOI={10.1140/epjc/s10052-021-09655-y},
   number={10},
   journal={The European Physical Journal C},
   publisher={Springer Science and Business Media LLC},
   author={Baxter, D. and Bloch, I. M. and Bodnia, E. and Chen, X. and Conrad, J. and Di Gangi, P. and Dobson, J. E. Y. and Durnford, D. and Haselschwardt, S. J. and Kaboth, A. and Lang, R. F. and Lin, Q. and Lippincott, W. H. and Liu, J. and Manalaysay, A. and McCabe, C. and Morå, K. D. and Naim, D. and Neilson, R. and Olcina, I. and Piro, M. -C. and Selvi, M. and von Krosigk, B. and Westerdale, S. and Yang, Y. and Zhou, N.},
   year={2021},
   month=oct }

@article{XENON1T,
  title = {Light Dark Matter Search with Ionization Signals in XENON1T},
  author = {Aprile, E. and Aalbers, J. and Agostini, F. and Alfonsi, M. and Althueser, L. and Amaro, F. D. and Antochi, V. C. and Angelino, E. and Arneodo, F. and Barge, D. and Baudis, L. and Bauermeister, B. and Bellagamba, L. and Benabderrahmane, M. L. and Berger, T. and Breur, P. A. and Brown, A. and Brown, E. and Bruenner, S. and Bruno, G. and Budnik, R. and Capelli, C. and Cardoso, J. M. R. and Cichon, D. and Coderre, D. and Colijn, A. P. and Conrad, J. and Cussonneau, J. P. and Decowski, M. P. and de Perio, P. and Depoian, A. and Di Gangi, P. and Di Giovanni, A. and Diglio, S. and Elykov, A. and Eurin, G. and Fei, J. and Ferella, A. D. and Fieguth, A. and Fulgione, W. and Gaemers, P. and Gallo Rosso, A. and Galloway, M. and Gao, F. and Garbini, M. and Grandi, L. and Greene, Z. and Hasterok, C. and Hils, C. and Hogenbirk, E. and Howlett, J. and Iacovacci, M. and Itay, R. and Joerg, F. and Kazama, S. and Kish, A. and Kobayashi, M. and Koltman, G. and Kopec, A. and Landsman, H. and Lang, R. F. and Levinson, L. and Lin, Q. and Lindemann, S. and Lindner, M. and Lombardi, F. and Lopes, J. A. M. and L\'opez Fune, E. and Macolino, C. and Mahlstedt, J. and Manfredini, A. and Marignetti, F. and Marrod\'an Undagoitia, T. and Masbou, J. and Mastroianni, S. and Messina, M. and Micheneau, K. and Miller, K. and Molinario, A. and Mor\aa{}, K. and Mosbacher, Y. and Murra, M. and Naganoma, J. and Ni, K. and Oberlack, U. and Odgers, K. and Palacio, J. and Pelssers, B. and Peres, R. and Pienaar, J. and Pizzella, V. and Plante, G. and Podviianiuk, R. and Qin, J. and Qiu, H. and Ram\'{\i}rez Garc\'{\i}a, D. and Reichard, S. and Riedel, B. and Rocchetti, A. and Rupp, N. and dos Santos, J. M. F. and Sartorelli, G. and \ifmmode \check{S}\else \v{S}\fi{}ar\ifmmode \check{c}\else \v{c}\fi{}evi\ifmmode \acute{c}\else \'{c}\fi{}, N. and Scheibelhut, M. and Schindler, S. and Schreiner, J. and Schulte, D. and Schumann, M. and Scotto Lavina, L. and Selvi, M. and Shagin, P. and Shockley, E. and Silva, M. and Simgen, H. and Therreau, C. and Thers, D. and Toschi, F. and Trinchero, G. and Tunnell, C. and Upole, N. and Vargas, M. and Volta, G. and Wack, O. and Wang, H. and Wei, Y. and Weinheimer, C. and Wenz, D. and Wittweg, C. and Wulf, J. and Ye, J. and Zhang, Y. and Zhu, T. and Zopounidis, J. P.},
  collaboration = {XENON Collaboration},
  journal = {Phys. Rev. Lett.},
  volume = {123},
  issue = {25},
  pages = {251801},
  numpages = {8},
  year = {2019},
  month = {Dec},
  publisher = {American Physical Society},
  doi = {10.1103/PhysRevLett.123.251801},
  url = {https://link.aps.org/doi/10.1103/PhysRevLett.123.251801}
}

@article{Akerib_2016,
   title={Improved Limits on Scattering of Weakly Interacting Massive Particles from Reanalysis of 2013 LUX Data},
   volume={116},
   ISSN={1079-7114},
   url={http://dx.doi.org/10.1103/PhysRevLett.116.161301},
   DOI={10.1103/physrevlett.116.161301},
   number={16},
   journal={Physical Review Letters},
   publisher={American Physical Society (APS)},
   author={Akerib, D. S. and Araújo, H. M. and Bai, X. and Bailey, A. J. and Balajthy, J. and Beltrame, P. and Bernard, E. P. and Bernstein, A. and Biesiadzinski, T. P. and Boulton, E. M. and Bradley, A. and Bramante, R. and Cahn, S. B. and Carmona-Benitez, M. C. and Chan, C. and Chapman, J. J. and Chiller, A. A. and Chiller, C. and Currie, A. and Cutter, J. E. and Davison, T. J. R. and de Viveiros, L. and Dobi, A. and Dobson, J. E. Y. and Druszkiewicz, E. and Edwards, B. N. and Faham, C. H. and Fiorucci, S. and Gaitskell, R. J. and Gehman, V. M. and Ghag, C. and Gibson, K. R. and Gilchriese, M. G. D. and Hall, C. R. and Hanhardt, M. and Haselschwardt, S. J. and Hertel, S. A. and Hogan, D. P. and Horn, M. and Huang, D. Q. and Ignarra, C. M. and Ihm, M. and Jacobsen, R. G. and Ji, W. and Kazkaz, K. and Khaitan, D. and Knoche, R. and Larsen, N. A. and Lee, C. and Lenardo, B. G. and Lesko, K. T. and Lindote, A. and Lopes, M. I. and Malling, D. C. and Manalaysay, A. and Mannino, R. L. and Marzioni, M. F. and McKinsey, D. N. and Mei, D.-M. and Mock, J. and Moongweluwan, M. and Morad, J. A. and Murphy, A. St. J. and Nehrkorn, C. and Nelson, H. N. and Neves, F. and O’Sullivan, K. and Oliver-Mallory, K. C. and Ott, R. A. and Palladino, K. J. and Pangilinan, M. and Pease, E. K. and Phelps, P. and Reichhart, L. and Rhyne, C. and Shaw, S. and Shutt, T. A. and Silva, C. and Solovov, V. N. and Sorensen, P. and Stephenson, S. and Sumner, T. J. and Szydagis, M. and Taylor, D. J. and Taylor, W. and Tennyson, B. P. and Terman, P. A. and Tiedt, D. R. and To, W. H. and Tripathi, M. and Tvrznikova, L. and Uvarov, S. and Verbus, J. R. and Webb, R. C. and White, J. T. and Whitis, T. J. and Witherell, M. S. and Wolfs, F. L. H. and Yazdani, K. and Young, S. K. and Zhang, C.},
   year={2016},
   month=apr }

@article{PandaX,
  title = {Dark Matter Search Results from $1.54\text{ }\text{ }\mathrm{Tonne}\ifmmode\cdot\else\textperiodcentered\fi{}\mathrm{Year}$ Exposure of PandaX-4T},
  author = {Bo, Zihao and Chen, Wei and Chen, Xun and Chen, Yunhua and Cheng, Zhaokan and Cui, Xiangyi and Fan, Yingjie and Fang, Deqing and Gao, Zhixing and Geng, Lisheng and Giboni, Karl and Guo, Xunan and Guo, Xuyuan and Guo, Zichao and Han, Chencheng and Han, Ke and He, Changda and He, Jinrong and Huang, Di and Huang, Houqi and Huang, Junting and Hou, Ruquan and Hou, Yu and Ji, Xiangdong and Ji, Xiangpan and Ju, Yonglin and Li, Chenxiang and Li, Jiafu and Li, Mingchuan and Li, Shuaijie and Li, Tao and Li, Zhiyuan and Lin, Qing and Liu, Jianglai and Lu, Congcong and Lu, Xiaoying and Luo, Lingyin and Luo, Yunyang and Ma, Wenbo and Ma, Yugang and Mao, Yajun and Meng, Yue and Ning, Xuyang and Pang, Binyu and Qi, Ningchun and Qian, Zhicheng and Ren, Xiangxiang and Shan, Dong and Shang, Xiaofeng and Shao, Xiyuan and Shen, Guofang and Shen, Manbin and Sun, Wenliang and Tao, Yi and Wang, Anqing and Wang, Guanbo and Wang, Hao and Wang, Jiamin and Wang, Lei and Wang, Meng and Wang, Qiuhong and Wang, Shaobo and Wang, Siguang and Wang, Wei and Wang, Xiuli and Wang, Xu and Wang, Zhou and Wei, Yuehuan and Wu, Weihao and Wu, Yuan and Xiao, Mengjiao and Xiao, Xiang and Xiong, Kaizhi and Xu, Yifan and Yao, Shunyu and Yan, Binbin and Yan, Xiyu and Yang, Yong and Ye, Peihua and Yu, Chunxu and Yuan, Ying and Yuan, Zhe and Yun, Youhui and Zeng, Xinning and Zhang, Minzhen and Zhang, Peng and Zhang, Shibo and Zhang, Shu and Zhang, Tao and Zhang, Wei and Zhang, Yang and Zhang, Yingxin and Zhang, Yuanyuan and Zhao, Li and Zhou, Jifang and Zhou, Jiaxu and Zhou, Jiayi and Zhou, Ning and Zhou, Xiaopeng and Zhou, Yubo and Zhou, Zhizhen},
  collaboration = {PandaX Collaboration},
  journal = {Phys. Rev. Lett.},
  volume = {134},
  issue = {1},
  pages = {011805},
  numpages = {8},
  year = {2025},
  month = {Jan},
  publisher = {American Physical Society},
  doi = {10.1103/PhysRevLett.134.011805},
  url = {https://link.aps.org/doi/10.1103/PhysRevLett.134.011805}
}

@article{LZ,
  title = {First Dark Matter Search Results from the LUX-ZEPLIN (LZ) Experiment},
  author = {Aalbers, J. and Akerib, D. S. and Akerlof, C. W. and Al Musalhi, A. K. and Alder, F. and Alqahtani, A. and Alsum, S. K. and Amarasinghe, C. S. and Ames, A. and Anderson, T. J. and Angelides, N. and Ara\'ujo, H. M. and Armstrong, J. E. and Arthurs, M. and Azadi, S. and Bailey, A. J. and Baker, A. and Balajthy, J. and Balashov, S. and Bang, J. and Bargemann, J. W. and Barry, M. J. and Barthel, J. and Bauer, D. and Baxter, A. and Beattie, K. and Belle, J. and Beltrame, P. and Bensinger, J. and Benson, T. and Bernard, E. P. and Bhatti, A. and Biekert, A. and Biesiadzinski, T. P. and Birch, H. J. and Birrittella, B. and Blockinger, G. M. and Boast, K. E. and Boxer, B. and Bramante, R. and Brew, C. A. J. and Br\'as, P. and Buckley, J. H. and Bugaev, V. V. and Burdin, S. and Busenitz, J. K. and Buuck, M. and Cabrita, R. and Carels, C. and Carlsmith, D. L. and Carlson, B. and Carmona-Benitez, M. C. and Cascella, M. and Chan, C. and Chawla, A. and Chen, H. and Cherwinka, J. J. and Chott, N. I. and Cole, A. and Coleman, J. and Converse, M. V. and Cottle, A. and Cox, G. and Craddock, W. W. and Creaner, O. and Curran, D. and Currie, A. and Cutter, J. E. and Dahl, C. E. and David, A. and Davis, J. and Davison, T. J. R. and Delgaudio, J. and Dey, S. and de Viveiros, L. and Dobi, A. and Dobson, J. E. Y. and Druszkiewicz, E. and Dushkin, A. and Edberg, T. K. and Edwards, W. R. and Elnimr, M. M. and Emmet, W. T. and Eriksen, S. R. and Faham, C. H. and Fan, A. and Fayer, S. and Fearon, N. M. and Fiorucci, S. and Flaecher, H. and Ford, P. and Francis, V. B. and Fraser, E. D. and Fruth, T. and Gaitskell, R. J. and Gantos, N. J. and Garcia, D. and Geffre, A. and Gehman, V. M. and Genovesi, J. and Ghag, C. and Gibbons, R. and Gibson, E. and Gilchriese, M. G. D. and Gokhale, S. and Gomber, B. and Green, J. and Greenall, A. and Greenwood, S. and van der Grinten, M. G. D. and Gwilliam, C. B. and Hall, C. R. and Hans, S. and Hanzel, K. and Harrison, A. and Hartigan-O'Connor, E. and Haselschwardt, S. J. and Hernandez, M. A. and Hertel, S. A. and Heuermann, G. and Hjemfelt, C. and Hoff, M. D. and Holtom, E. and Hor, J. Y-K. and Horn, M. and Huang, D. Q. and Hunt, D. and Ignarra, C. M. and Jacobsen, R. G. and Jahangir, O. and James, R. S. and Jeffery, S. N. and Ji, W. and Johnson, J. and Kaboth, A. C. and Kamaha, A. C. and Kamdin, K. and Kasey, V. and Kazkaz, K. and Keefner, J. and Khaitan, D. and Khaleeq, M. and Khazov, A. and Khurana, I. and Kim, Y. D. and Kocher, C. D. and Kodroff, D. and Korley, L. and Korolkova, E. V. and Kras, J. and Kraus, H. and Kravitz, S. and Krebs, H. J. and Kreczko, L. and Krikler, B. and Kudryavtsev, V. A. and Kyre, S. and Landerud, B. and Leason, E. A. and Lee, C. and Lee, J. and Leonard, D. S. and Leonard, R. and Lesko, K. T. and Levy, C. and Li, J. and Liao, F.-T. and Liao, J. and Lin, J. and Lindote, A. and Linehan, R. and Lippincott, W. H. and Liu, R. and Liu, X. and Liu, Y. and Loniewski, C. and Lopes, M. I. and Lopez Asamar, E. and L\'opez Paredes, B. and Lorenzon, W. and Lucero, D. and Luitz, S. and Lyle, J. M. and Majewski, P. A. and Makkinje, J. and Malling, D. C. and Manalaysay, A. and Manenti, L. and Mannino, R. L. and Marangou, N. and Marzioni, M. F. and Maupin, C. and McCarthy, M. E. and McConnell, C. T. and McKinsey, D. N. and McLaughlin, J. and Meng, Y. and Migneault, J. and Miller, E. H. and Mizrachi, E. and Mock, J. A. and Monte, A. and Monzani, M. E. and Morad, J. A. and Morales Mendoza, J. D. and Morrison, E. and Mount, B. J. and Murdy, M. and Murphy, A. St. J. and Naim, D. and Naylor, A. and Nedlik, C. and Nehrkorn, C. and Neves, F. and Nguyen, A. and Nikoleyczik, J. A. and Nilima, A. and O'Dell, J. and O'Neill, F. G. and O'Sullivan, K. and Olcina, I. and Olevitch, M. A. and Oliver-Mallory, K. C. and Orpwood, J. and Pagenkopf, D. and Pal, S. and Palladino, K. J. and Palmer, J. and Pangilinan, M. and Parveen, N. and Patton, S. J. and Pease, E. K. and Penning, B. and Pereira, C. and Pereira, G. and Perry, E. and Pershing, T. and Peterson, I. B. and Piepke, A. and Podczerwinski, J. and Porzio, D. and Powell, S. and Preece, R. M. and Pushkin, K. and Qie, Y. and Ratcliff, B. N. and Reichenbacher, J. and Reichhart, L. and Rhyne, C. A. and Richards, A. and Riffard, Q. and Rischbieter, G. R. C. and Rodrigues, J. P. and Rodriguez, A. and Rose, H. J. and Rosero, R. and Rossiter, P. and Rushton, T. and Rutherford, G. and Rynders, D. and Saba, J. S. and Santone, D. and Sazzad, A. B. M. R. and Schnee, R. W. and Scovell, P. R. and Seymour, D. and Shaw, S. and Shutt, T. and Silk, J. J. and Silva, C. and Sinev, G. and Skarpaas, K. and Skulski, W. and Smith, R. and Solmaz, M. and Solovov, V. N. and Sorensen, P. and Soria, J. and Stancu, I. and Stark, M. R. and Stevens, A. and Stiegler, T. M. and Stifter, K. and Studley, R. and Suerfu, B. and Sumner, T. J. and Sutcliffe, P. and Swanson, N. and Szydagis, M. and Tan, M. and Taylor, D. J. and Taylor, R. and Taylor, W. C. and Temples, D. J. and Tennyson, B. P. and Terman, P. A. and Thomas, K. J. and Tiedt, D. R. and Timalsina, M. and To, W. H. and Tom\'as, A. and Tong, Z. and Tovey, D. R. and Tranter, J. and Trask, M. and Tripathi, M. and Tronstad, D. R. and Tull, C. E. and Turner, W. and Tvrznikova, L. and Utku, U. and Va'vra, J. and Vacheret, A. and Vaitkus, A. C. and Verbus, J. R. and Voirin, E. and Waldron, W. L. and Wang, A. and Wang, B. and Wang, J. J. and Wang, W. and Wang, Y. and Watson, J. R. and Webb, R. C. and White, A. and White, D. T. and White, J. T. and White, R. G. and Whitis, T. J. and Williams, M. and Wisniewski, W. J. and Witherell, M. S. and Wolfs, F. L. H. and Wolfs, J. D. and Woodford, S. and Woodward, D. and Worm, S. D. and Wright, C. J. and Xia, Q. and Xiang, X. and Xiao, Q. and Xu, J. and Yeh, M. and Yin, J. and Young, I. and Zarzhitsky, P. and Zuckerman, A. and Zweig, E. A.},
  collaboration = {LUX-ZEPLIN Collaboration},
  journal = {Phys. Rev. Lett.},
  volume = {131},
  issue = {4},
  pages = {041002},
  numpages = {11},
  year = {2023},
  month = {Jul},
  publisher = {American Physical Society},
  doi = {10.1103/PhysRevLett.131.041002},
  url = {https://link.aps.org/doi/10.1103/PhysRevLett.131.041002}
}

@misc{XENONnt,
      title={WIMP Dark Matter Search using a 3.1 tonne $\times$ year Exposure of the XENONnT Experiment}, 
      author={E. Aprile and J. Aalbers and K. Abe and S. Ahmed Maouloud and L. Althueser and B. Andrieu and E. Angelino and D. Antón Martin and S. R. Armbruster and F. Arneodo and L. Baudis and M. Bazyk and L. Bellagamba and R. Biondi and A. Bismark and K. Boese and A. Brown and G. Bruno and R. Budnik and C. Cai and C. Capelli and J. M. R. Cardoso and A. P. Cimental Chávez and A. P. Colijn and J. Conrad and J. J. Cuenca-García and V. D'Andrea and L. C. Daniel Garcia and M. P. Decowski and A. Deisting and C. Di Donato and P. Di Gangi and S. Diglio and K. Eitel and S. el Morabit and A. Elykov and A. D. Ferella and C. Ferrari and H. Fischer and T. Flehmke and M. Flierman and D. Fuchs and W. Fulgione and C. Fuselli and P. Gaemers and R. Gaior and F. Gao and S. Ghosh and R. Giacomobono and F. Girard and R. Glade-Beucke and L. Grandi and J. Grigat and H. Guan and M. Guida and P. Gyorgy and R. Hammann and A. Higuera and C. Hils and L. Hoetzsch and N. F. Hood and M. Iacovacci and Y. Itow and J. Jakob and F. Joerg and Y. Kaminaga and M. Kara and P. Kavrigin and S. Kazama and P. Kharbanda and M. Kobayashi and D. Koke and K. Kooshkjalali and A. Kopec and H. Landsman and R. F. Lang and L. Levinson and I. Li and S. Li and S. Liang and Z. Liang and Y. -T. Lin and S. Lindemann and M. Lindner and K. Liu and M. Liu and J. Loizeau and F. Lombardi and J. Long and J. A. M. Lopes and G. M. Lucchetti and T. Luce and Y. Ma and C. Macolino and J. Mahlstedt and A. Mancuso and L. Manenti and F. Marignetti and T. Marrodán Undagoitia and K. Martens and J. Masbou and S. Mastroianni and A. Melchiorre and J. Merz and M. Messina and A. Michael and K. Miuchi and A. Molinario and S. Moriyama and K. Morå and Y. Mosbacher and M. Murra and J. Müller and K. Ni and U. Oberlack and B. Paetsch and Y. Pan and Q. Pellegrini and R. Peres and C. Peters and J. Pienaar and M. Pierre and G. Plante and T. R. Pollmann and L. Principe and J. Qi and J. Qin and D. Ramírez García and M. Rajado and A. Ravindran and A. Razeto and R. Singh and L. Sanchez and J. M. F. dos Santos and I. Sarnoff and G. Sartorelli and J. Schreiner and P. Schulte and H. Schulze Eißing and M. Schumann and L. Scotto Lavina and M. Selvi and F. Semeria and P. Shagin and S. Shi and J. Shi and M. Silva and H. Simgen and A. Stevens and C. Szyszka and A. Takeda and Y. Takeuchi and P. -L. Tan and D. Thers and G. Trinchero and C. D. Tunnell and F. Tönnies and K. Valerius and S. Vecchi and S. Vetter and F. I. Villazon Solar and G. Volta and C. Weinheimer and M. Weiss and D. Wenz and C. Wittweg and V. H. S. Wu and Y. Xing and D. Xu and Z. Xu and M. Yamashita and J. Yang and L. Yang and J. Ye and L. Yuan and G. Zavattini and Y. Zhao and M. Zhong},
      year={2025},
      eprint={2502.18005},
      archivePrefix={arXiv},
      primaryClass={hep-ex},
      url={https://arxiv.org/abs/2502.18005}, 
}

@article{COSINE-100:2024nfa,
    author = "Carlin, Nelson and others",
    collaboration = "COSINE-100",
    title = "{COSINE-100 full dataset challenges the annual modulation signal of DAMA/LIBRA}",
    eprint = "2409.13226",
    archivePrefix = "arXiv",
    primaryClass = "hep-ex",
    doi = "10.1126/sciadv.adv6503",
    journal = "Sci. Adv.",
    volume = "11",
    number = "36",
    pages = "adv6503",
    year = "2025"
}

@article{Abdullah:2022zue,
    author = "Abdullah, M. and others",
    title = "{Coherent elastic neutrino-nucleus scattering: Terrestrial and astrophysical applications}",
    eprint = "2203.07361",
    archivePrefix = "arXiv",
    primaryClass = "hep-ph",
    doi = "10.2172/1856010",
    month = "3",
    year = "2022"
}

@article{Freedman,
  title = {Coherent effects of a weak neutral current},
  author = {Freedman, Daniel Z.},
  journal = {Phys. Rev. D},
  volume = {9},
  issue = {5},
  pages = {1389--1392},
  numpages = {0},
  year = {1974},
  month = {Mar},
  publisher = {American Physical Society},
  doi = {10.1103/PhysRevD.9.1389},
  url = {https://link.aps.org/doi/10.1103/PhysRevD.9.1389}
}

@article{Akimov_2017,
   title={Observation of coherent elastic neutrino-nucleus scattering},
   volume={357},
   ISSN={1095-9203},
   url={http://dx.doi.org/10.1126/science.aao0990},
   DOI={10.1126/science.aao0990},
   number={6356},
   journal={Science},
   publisher={American Association for the Advancement of Science (AAAS)},
   author={Akimov, D. and Albert, J. B. and An, P. and Awe, C. and Barbeau, P. S. and Becker, B. and Belov, V. and Brown, A. and Bolozdynya, A. and Cabrera-Palmer, B. and Cervantes, M. and Collar, J. I. and Cooper, R. J. and Cooper, R. L. and Cuesta, C. and Dean, D. J. and Detwiler, J. A. and Eberhardt, A. and Efremenko, Y. and Elliott, S. R. and Erkela, E. M. and Fabris, L. and Febbraro, M. and Fields, N. E. and Fox, W. and Fu, Z. and Galindo-Uribarri, A. and Green, M. P. and Hai, M. and Heath, M. R. and Hedges, S. and Hornback, D. and Hossbach, T. W. and Iverson, E. B. and Kaufman, L. J. and Ki, S. and Klein, S. R. and Khromov, A. and Konovalov, A. and Kremer, M. and Kumpan, A. and Leadbetter, C. and Li, L. and Lu, W. and Mann, K. and Markoff, D. M. and Miller, K. and Moreno, H. and Mueller, P. E. and Newby, J. and Orrell, J. L. and Overman, C. T. and Parno, D. S. and Penttila, S. and Perumpilly, G. and Ray, H. and Raybern, J. and Reyna, D. and Rich, G. C. and Rimal, D. and Rudik, D. and Scholberg, K. and Scholz, B. J. and Sinev, G. and Snow, W. M. and Sosnovtsev, V. and Shakirov, A. and Suchyta, S. and Suh, B. and Tayloe, R. and Thornton, R. T. and Tolstukhin, I. and Vanderwerp, J. and Varner, R. L. and Virtue, C. J. and Wan, Z. and Yoo, J. and Yu, C.-H. and Zawada, A. and Zettlemoyer, J. and Zderic, A. M.},
   year={2017},
   month=sep, pages={1123–1126} }

@article{O_Hare_2021,
   title={New Definition of the Neutrino Floor for Direct Dark Matter Searches},
   volume={127},
   ISSN={1079-7114},
   url={http://dx.doi.org/10.1103/PhysRevLett.127.251802},
   DOI={10.1103/physrevlett.127.251802},
   number={25},
   journal={Physical Review Letters},
   publisher={American Physical Society (APS)},
   author={O’Hare, Ciaran A. J.},
   year={2021},
   month=dec }

@article{Feng_2011,
   title={Isospin-violating dark matter},
   volume={703},
   ISSN={0370-2693},
   url={http://dx.doi.org/10.1016/j.physletb.2011.07.083},
   DOI={10.1016/j.physletb.2011.07.083},
   number={2},
   journal={Physics Letters B},
   publisher={Elsevier BV},
   author={Feng, Jonathan L. and Kumar, Jason and Marfatia, Danny and Sanford, David},
   year={2011},
   month=sep, pages={124–127} }

@article{Feng_2013,
   title={Xenophobic dark matter},
   volume={88},
   ISSN={1550-2368},
   url={http://dx.doi.org/10.1103/PhysRevD.88.015021},
   DOI={10.1103/physrevd.88.015021},
   number={1},
   journal={Physical Review D},
   publisher={American Physical Society (APS)},
   author={Feng, Jonathan L. and Kumar, Jason and Sanford, David},
   year={2013},
   month=jul }

@article{PandaxSolar,
  title = {First Indication of Solar $^{8}\mathrm{B}$ Neutrinos through Coherent Elastic Neutrino-Nucleus Scattering in PandaX-4T},
  author = {Bo, Zihao and Chen, Wei and Chen, Xun and Chen, Yunhua and Cheng, Zhaokan and Cui, Xiangyi and Fan, Yingjie and Fang, Deqing and Gao, Zhixing and Geng, Lisheng and Giboni, Karl and Guo, Xunan and Guo, Xuyuan and Guo, Zichao and Han, Chencheng and Han, Ke and He, Changda and He, Jinrong and Huang, Di and Huang, Houqi and Huang, Junting and Hou, Ruquan and Hou, Yu and Ji, Xiangdong and Ji, Xiangpan and Ju, Yonglin and Li, Chenxiang and Li, Jiafu and Li, Mingchuan and Li, Shuaijie and Li, Tao and Li, Zhiyuan and Lin, Qing and Liu, Jianglai and Lu, Congcong and Lu, Xiaoying and Luo, Lingyin and Luo, Yunyang and Ma, Wenbo and Ma, Yugang and Mao, Yajun and Meng, Yue and Ning, Xuyang and Pang, Binyu and Qi, Ningchun and Qian, Zhicheng and Ren, Xiangxiang and Shan, Dong and Shang, Xiaofeng and Shao, Xiyuan and Shen, Guofang and Shen, Manbin and Sun, Wenliang and Tao, Yi and Wang, Anqing and Wang, Guanbo and Wang, Hao and Wang, Jiamin and Wang, Lei and Wang, Meng and Wang, Qiuhong and Wang, Shaobo and Wang, Siguang and Wang, Wei and Wang, Xiuli and Wang, Xu and Wang, Zhou and Wei, Yuehuan and Wu, Weihao and Wu, Yuan and Xiao, Mengjiao and Xiao, Xiang and Xiong, Kaizhi and Xu, Yifan and Yao, Shunyu and Yan, Binbin and Yan, Xiyu and Yang, Yong and Ye, Peihua and Yu, Chunxu and Yuan, Ying and Yuan, Zhe and Yun, Youhui and Zeng, Xinning and Zhang, Minzhen and Zhang, Peng and Zhang, Shibo and Zhang, Shu and Zhang, Tao and Zhang, Wei and Zhang, Yang and Zhang, Yingxin and Zhang, Yuanyuan and Zhao, Li and Zhou, Jifang and Zhou, Jiaxu and Zhou, Jiayi and Zhou, Ning and Zhou, Xiaopeng and Zhou, Yubo and Zhou, Zhizhen},
  collaboration = {PandaX Collaboration},
  journal = {Phys. Rev. Lett.},
  volume = {133},
  issue = {19},
  pages = {191001},
  numpages = {9},
  year = {2024},
  month = {Nov},
  publisher = {American Physical Society},
  doi = {10.1103/PhysRevLett.133.191001},
  url = {https://link.aps.org/doi/10.1103/PhysRevLett.133.191001}
}

@article{XenonntSolar,
  title = {First Indication of Solar $^{8}\mathrm{B}$ Neutrinos via Coherent Elastic Neutrino-Nucleus Scattering with XENONnT},
  author = {Aprile, E. and Aalbers, J. and Abe, K. and Ahmed Maouloud, S. and Althueser, L. and Andrieu, B. and Angelino, E. and Ant\'on Martin, D. and Arneodo, F. and Baudis, L. and Bazyk, M. and Bellagamba, L. and Biondi, R. and Bismark, A. and Boese, K. and Brown, A. and Bruno, G. and Budnik, R. and Cai, C. and Capelli, C. and Cardoso, J. M. R. and Cimental Ch\'avez, A. P. and Colijn, A. P. and Conrad, J. and Cuenca-Garc\'{\i}a, J. J. and D'Andrea, V. and Daniel Garcia, L. C. and Decowski, M. P. and Deisting, A. and Di Donato, C. and Di Gangi, P. and Diglio, S. and Eitel, K. and Elykov, A. and Ferella, A. D. and Ferrari, C. and Fischer, H. and Flehmke, T. and Flierman, M. and Fulgione, W. and Fuselli, C. and Gaemers, P. and Gaior, R. and Galloway, M. and Gao, F. and Ghosh, S. and Giacomobono, R. and Glade-Beucke, R. and Grandi, L. and Grigat, J. and Guan, H. and Guida, M. and Gyorgy, P. and Hammann, R. and Higuera, A. and Hils, C. and Hoetzsch, L. and Hood, N. F. and Iacovacci, M. and Itow, Y. and Jakob, J. and Joerg, F. and Kaminaga, Y. and Kara, M. and Kavrigin, P. and Kazama, S. and Kobayashi, M. and Koke, D. and Kopec, A. and Kuger, F. and Landsman, H. and Lang, R. F. and Levinson, L. and Li, I. and Li, S. and Liang, S. and Lin, Y.-T. and Lindemann, S. and Lindner, M. and Liu, K. and Liu, M. and Loizeau, J. and Lombardi, F. and Long, J. and Lopes, J. A. M. and Luce, T. and Ma, Y. and Macolino, C. and Mahlstedt, J. and Mancuso, A. and Manenti, L. and Marignetti, F. and Marrod\'an Undagoitia, T. and Martens, K. and Masbou, J. and Masson, E. and Mastroianni, S. and Melchiorre, A. and Merz, J. and Messina, M. and Michael, A. and Miuchi, K. and Molinario, A. and Moriyama, S. and Mor\aa{}, K. and Mosbacher, Y. and Murra, M. and M\"uller, J. and Ni, K. and Oberlack, U. and Paetsch, B. and Pan, Y. and Pellegrini, Q. and Peres, R. and Peters, C. and Pienaar, J. and Pierre, M. and Plante, G. and Pollmann, T. R. and Principe, L. and Qi, J. and Qin, J. and Ram\'{\i}rez Garc\'{\i}a, D. and Rajado, M. and Singh, R. and Sanchez, L. and dos Santos, J. M. F. and Sarnoff, I. and Sartorelli, G. and Schreiner, J. and Schulte, P. and Schulze Ei\ss{}ing, H. and Schumann, M. and Scotto Lavina, L. and Selvi, M. and Semeria, F. and Shagin, P. and Shi, S. and Shi, J. and Silva, M. and Simgen, H. and Takeda, A. and Tan, P.-L. and Thers, D. and Toschi, F. and Trinchero, G. and Tunnell, C. D. and T\"onnies, F. and Valerius, K. and Vecchi, S. and Vetter, S. and Villazon Solar, F. I. and Volta, G. and Weinheimer, C. and Weiss, M. and Wenz, D. and Wittweg, C. and Wu, V. H. S. and Xing, Y. and Xu, D. and Xu, Z. and Yamashita, M. and Yang, L. and Ye, J. and Yuan, L. and Zavattini, G. and Zhong, M.},
  collaboration = {XENON Collaboration},
  journal = {Phys. Rev. Lett.},
  volume = {133},
  issue = {19},
  pages = {191002},
  numpages = {11},
  year = {2024},
  month = {Nov},
  publisher = {American Physical Society},
  doi = {10.1103/PhysRevLett.133.191002},
  url = {https://link.aps.org/doi/10.1103/PhysRevLett.133.191002}
}

@misc{LZSolar,
      title={Searches for Light Dark Matter and Evidence of Coherent Elastic Neutrino-Nucleus Scattering of Solar Neutrinos with the LUX-ZEPLIN (LZ) Experiment}, 
      author={D. S. Akerib and A. K. Al Musalhi and F. Alder and B. J. Almquist and C. S. Amarasinghe and A. Ames and T. J. Anderson and N. Angelides and H. M. Araújo and J. E. Armstrong and M. Arthurs and A. Baker and S. Balashov and J. Bang and J. W. Bargemann and E. E. Barillier and D. Bauer and K. Beattie and A. Bhatti and T. P. Biesiadzinski and H. J. Birch and E. Bishop and G. M. Blockinger and C. A. J. Brew and P. Brás and S. Burdin and M. C. Carmona-Benitez and M. Carter and A. Chawla and H. Chen and Y. T. Chin and N. I. Chott and S. Contreras and M. V. Converse and R. Coronel and A. Cottle and G. Cox and D. Curran and C. E. Dahl and I. Darlington and S. Dave and A. David and J. Delgaudio and S. Dey and L. de Viveiros and L. Di Felice and C. Ding and J. E. Y. Dobson and E. Druszkiewicz and S. Dubey and C. L. Dunbar and S. R. Eriksen and S. Fayer and N. M. Fearon and N. Fieldhouse and S. Fiorucci and H. Flaecher and E. D. Fraser and T. M. A. Fruth and P. W. Gaemers and R. J. Gaitskell and A. Geffre and J. Genovesi and C. Ghag and J. Ghamsari and A. Ghosh and S. Ghosh and R. Gibbons and S. Gokhale and J. Green and M. G. D. van der Grinten and J. J. Haiston and C. R. Hall and T. Hall and R. N Hampp and S. J. Haselschwardt and M. A. Hernandez and S. A. Hertel and G. J. Homenides and M. Horn and D. Q. Huang and D. Hunt and E. Jacquet and R. James and K. Jenkins and A. C. Kaboth and A. C. Kamaha and M. K. Kannichankandy and D. Khaitan and A. Khazov and J. Kim and Y. D. Kim and D. Kodroff and E. V. Korolkova and H. Kraus and S. Kravitz and L. Kreczko and V. A. Kudryavtsev and C. Lawes and D. S. Leonard and K. T. Lesko and C. Levy and J. Lin and A. Lindote and W. H. Lippincott and J. Long and M. I. Lopes and W. Lorenzon and C. Lu and S. Luitz and W. Ma and V. Mahajan and P. A. Majewski and A. Manalaysay and R. L. Mannino and R. J. Matheson and C. Maupin and M. E. McCarthy and D. N. McKinsey and J. McLaughlin and J. B. McLaughlin and R. McMonigle and B. Mitra and E. Mizrachi and M. E. Monzani and K. Moraa and E. Morrison and B. J. Mount and M. Murdy and A. St. J. Murphy and H. N. Nelson and F. Neves and A. Nguyen and C. L. O'Brien and F. H. O'Shea and I. Olcina and K. C. Oliver-Mallory and J. Orpwood and K. Y Oyulmaz and K. J. Palladino and N. J. Pannifer and N. Parveen and S. J. Patton and B. Penning and G. Pereira and E. Perry and T. Pershing and A. Piepke and S. S. Poudel and Y. Qie and J. Reichenbacher and C. A. Rhyne and G. R. C. Rischbieter and E. Ritchey and H. S. Riyat and R. Rosero and N. J. Rowe and T. Rushton and D. Rynders and S. Saltão and D. Santone and I. Sargeant and A. B. M. R. Sazzad and R. W. Schnee and G. Sehr and B. Shafer and S. Shaw and W. Sherman and K. Shi and T. Shutt and C. Silva and G. Sinev and J. Siniscalco and A. M. Slivar and R. Smith and V. N. Solovov and P. Sorensen and J. Soria and T. J. Sumner and A. Swain and M. Szydagis and D. R. Tiedt and M. Timalsina and D. R. Tovey and J. Tranter and M. Trask and K. Trengove and M. Tripathi and A. Usón and A. C. Vaitkus and O. Valentino and V. Velan and A. Wang and J. J. Wang and Y. Wang and L. Weeldreyer and T. J. Whitis and K. Wild and M. Williams and J. Winnicki and L. Wolf and F. L. H. Wolfs and S. Woodford and D. Woodward and C. J. Wright and Q. Xia and J. Xu and Y. Xu and M. Yeh and D. Yeum and J. Young and W. Zha and H. Zhang and T. Zhang and Y. Zhou},
      year={2025},
      eprint={2512.08065},
      archivePrefix={arXiv},
      primaryClass={hep-ex},
      url={https://arxiv.org/abs/2512.08065}, 
}

@misc{deromeri2025,
      title={New light mediators and the neutrino fog: Implications from XENONnT nuclear recoil data}, 
      author={Valentina De Romeri and Anirban Majumdar and Dimitrios K. Papoulias and Rahul Srivastava},
      year={2025},
      eprint={2512.08853},
      archivePrefix={arXiv},
      primaryClass={hep-ph},
      url={https://arxiv.org/abs/2512.08853}, 
}

@article{DeRomeri2023,
   title={Physics implications of a combined analysis of COHERENT CsI and LAr data},
   volume={2023},
   ISSN={1029-8479},
   url={http://dx.doi.org/10.1007/JHEP04(2023)035},
   DOI={10.1007/jhep04(2023)035},
   number={4},
   journal={Journal of High Energy Physics},
   publisher={Springer Science and Business Media LLC},
   author={De Romeri, V. and Miranda, O. G. and Papoulias, D. K. and Sanchez Garcia, G. and Tórtola, M. and Valle, J. W. F.},
   year={2023},
   month=apr }

\end{document}